%Paper: hep-ph/9309215
%From: yoshida@gauge.scphys.kyoto-u.ac.jp
%Date: Thu, 02 Sep 93 18:15:40 +0900
%Date (revised): Thu, 23 Dec 93 01:58:32 +0900

%%%%%%%%%%%   READ ME  READ ME  READ ME  READ ME  READ ME   %%%%%
%                                                               %
%    This paper has nine figures appended in a second part      %
%    as a uuencoded compressed tar file with instructions for   %
%    unpacking (Please search for '#!').                        %
%                                                               %
%%%%%%%%%%%%%%%%%%%%%%%%%%%%%%%%%%%%%%%%%%%%%%%%%%%%%%%%%%%%%%%%%

\input phyzzx
\input epsf
%\physrev

%%%%%%%%%%%%%%%%%%%%%%%%%%%%%%%%%%%%%%%%%%%%%%%%%%%%%%%%%%%%%%%%%
%%%%%%%%%%%%%%%%%%%%  MACRO %%%%%%%%%%%%%%%%%%%%%%%%%%%%%%%%%%%%%
%%%%%%%%%%%%%%%%%%%%%%%%%%%%%%%%%%%%%%%%%%%%%%%%%%%%%%%%%%%%%%%%%
\catcode`@=11
%
%%%%%%%%%%%%%% Macros for the Title Page
\newtoks\KUNS
\newtoks\HETH
\newtoks\monthyear
\Pubnum={KUNS~\the\KUNS\cr HE(TH)~\the\HETH\cr hep-ph/9309215}
\monthyear={\monthname,\ \number\year}
\def\p@bblock{\begingroup \tabskip=\hsize minus \hsize
   \baselineskip=1.5\ht\strutbox \topspace-2\baselineskip
   \halign to\hsize{\strut ##\hfil\tabskip=0pt\crcr
   \the\Pubnum\cr \the\monthyear\cr }\endgroup}
\def\bftitlestyle#1{\par\begingroup \titleparagraphs
     \iftwelv@\fourteenpoint\else\twelvepoint\fi
   \noindent {\bf #1}\par\endgroup }
\def\title#1{\vskip\frontpageskip \bftitlestyle{#1} \vskip\headskip }
\def\Kyoto{\address{Department of Physics,~Kyoto University \break
                            Kyoto~606,~JAPAN}}

%%%%%%%%%%%%%% Style of page numbers
\paperfootline={\hss\iffrontpage\else\ifp@genum%
                \tenrm --\thinspace\folio\thinspace --\hss\fi\fi}
\footline=\paperfootline
%%%%%%%%%%%%%% Macros for References
% Let's make \refmark larger

%
%Redefinition of \journal and new macros
\def\journal#1&#2(#3){\begingroup \let\journal=\dummyj@urnal
    \unskip, \sl #1\unskip~\bf\ignorespaces #2\rm
    (\afterassignment\j@ur \count255=#3) \endgroup\ignorespaces }
\def\andjournal#1&#2(#3){\begingroup \let\journal=\dummyj@urnal
    \sl #1\unskip~\bf\ignorespaces #2\rm
    (\afterassignment\j@ur \count255=#3) \endgroup\ignorespaces }
\def\andvol&#1(#2){\begingroup \let\journal=\dummyj@urnal
    \bf\ignorespaces #1\rm
    (\afterassignment\j@ur \count255=#2) \endgroup\ignorespaces }
\def\NP{Nucl.~Phys. }
\def\PR{Phys.~Rev. }
\def\PRL{Phys.~Rev.~Lett. }
\def\PL{Phys.~Lett. }
\def\PTP{Prog.~Theor.~Phys. }
\def\CMP{Commun.~Math.~Phys.}

%%%%%%%%%%%%%% Macro for Acknowledgements

%%%%%%%%%%%%%% Macro for math formulas
\def\overbar#1{\vbox{\ialign{##\crcr
          \hskip 1.5pt\hrulefill\hskip 1.1pt
          \crcr\noalign{\kern-1pt\vskip0.125cm\nointerlineskip}
          $\hfil\displaystyle{#1}\hfil$\crcr}}}
\def\figmark#1{\par\vskip0.5cm{\hbox{\centerline{
           \vbox{\hrule height0.5pt
           \hbox{\vrule width0.5pt\hskip4pt
           \vbox{\vskip5pt\hbox{Fig.#1}\vskip4pt}
           \hskip4pt\vrule width0.5pt}
           \hrule height0.5pt}}}}\par\noindent\hskip0.1cm\hskip-0.1cm}

\catcode`@=12
%%%%%%%%%%%%%%%%%%%%%%%%%%%%%%%%%%%%%%%%%%%%%%%%%%%%%%%%%%%%%%%%%
%%%%%%%%%%%%%%%%%%%%%%%% DEFINITIONS %%%%%%%%%%%%%%%%%%%%%%%%%%%%
%%%%%%%%%%%%%%%%%%%%%%%%%%%%%%%%%%%%%%%%%%%%%%%%%%%%%%%%%%%%%%%%%
\def\1#1{{1 \over {#1}}}
\def\2#1{{{#1} \over 2}}
\def\3#1{{{#1} \over 3}}
\def\4#1{{{#1} \over 4}}
\def\l{\ell}
\def\rp{\varphi}
\def\grp{g^2{\rp}^2}
\def\p{\partial}
\def\widebar#1{
	\vbox{\ialign{##\crcr
	\hskip 1.5pt\hrulefill\hskip 1.1pt
	\crcr\noalign{\kern-1pt\vskip0.07cm\nointerlineskip}
	$\hfil\displaystyle{#1}\hfil$\crcr}}}
\def\sss{\scriptscriptstyle}

%%%%%%%%%%%%%%%     rm or cal characters    %%%%%%%%%%%%%%%%
\def\F{{\rm F}}
\def\B{{\rm B}}

\def\MS{{\widebar {\rm MS}}}

\def\MDB{{\widebar {\rm MD}}}
\def\MDS{{\wb {\rm {\sss MD}}}}
\def\MSS{{\wb {\rm {\sss MS}}}}
%%%%%%%%%%%%%%%     constants	   %%%%%%%%%%%%%%%%%
\def\gpai{{g^2 \over 16{\pi}^2}}

%%%%%%%%%%%%%%%%%%%     mass scale     %%%%%%%%%%%%%%%%%%%%%
\def\m{\mu^2}
\def\momu{{m^2 \over \m }}
\def\muom{{\m  \over m^2}}
\def\lnmzm{ \ln{{\mu}_0^2 \over m^2} }
\def\lnmzp{ \ln{{\mu}_0^2 \over -p^2} }
\def\MF{M^2_{ {\rm {\sss F}} }}
\def\MB{M^2_{ {\rm {\sss B}} }}
\def\MFF{M^4_{ {\rm {\sss F}} }}
\def\MBF{M^4_{ {\rm {\sss B}} }}

\def\sF{s_{{\sss \F}}}
\def\sB{s_{{\sss \B}}}
\def\sFB{\widebar s_{{\sss \F}}}
\def\sBB{\widebar s_{{\sss \B}}}
\def\smB{{\widebar s}_m}
\def\stF{{\widetilde s}_{{\sss \F}}}

\def\ks{k\kern-6pt\hbox{$ /$}}
\def\ps{p\kern-5pt\hbox{$ /$}}
\def\qs{q\kern-5pt\hbox{$ /$}}
\def\rs{r\kern-5pt\hbox{$ /$}}
%%%%%%%%%%%%%%%%%%%     differential     %%%%%%%%%%%%%%%%%%%
\def\p{\partial}
\def\ds{{\p}\kern-6pt\hbox{$ /$}}
\def\d{ \m {d \over d \m} }
\def\dmu{ \mu {d \over d \mu} }
\def\der#1{{\partial \over \partial #1}}
%%%%%%%%%%%%%%%%%%%     bar or tilde    %%%%%%%%%%%%%%%%%%%%
\def\wt{\widetilde}
\def\wb{\widebar}
\def\ebar{{\widebar {\,\epsilon\,}}}

%%%%%%%%%%%%%%%%%%%     other definitions    %%%%%%%%%%%%%%%

\def\LINT{ \int {d^d k \over (2\pi)^d i} }
\def\simg{{\raise3pt\hbox{$>$}}\kern-9pt\lower3pt\hbox{$\sim$}}
\def\simg{\gsim}

\def\ggll{{\raise3pt\hbox{$>$}}\kern-8pt\lower3pt\hbox{$<$}}
\def\llgg{{\raise3pt\hbox{$<$}}\kern-8pt\lower3pt\hbox{$>$}}

\def\VLOOP{V_{{\rm loop}}}
\def\draw#1{\vbox{#1}}
%%%%%%%%%%%%%%%%%%%%%%%%%%%%%%%%%%%%%%%%%%%%%%%%%%%%%%%%%%%%%%%%%
%%%%%%%%%%%%%%%%%%%%%%%%% REFERENCES %%%%%%%%%%%%%%%%%%%%%%%%%%%%
%%%%%%%%%%%%%%%%%%%%%%%%%%%%%%%%%%%%%%%%%%%%%%%%%%%%%%%%%%%%%%%%%
\REF\GQW{
	H.~Georgi, H.~Quinn and S.~Weinberg
		\journal \PRL &33 (74) 451.}
\REF\CW{
	S.~Coleman and E.~Weinberg
		\journal \PR &D7 (73) 1888.}
\REF\KASTENING{
	B.~Kastening
	\journal \PL &B283 (92) 287.}
\REF\EPi{
	M.~Bando, T.~Kugo, N.~Maekawa and H.~Nakano
	\journal \PL &B301 (93) 81.}
\REF\FJSE{
	C.~Ford, D.~R.~T.~Jones, P.~W.~Stephenson and M.~B.~Einhorn
	\journal \NP  \break &B395 (93) 17.}
\REF\SHER{
	See for instance;
	M.~Sher, {\it Phys.~Rep.} {\bf 179} (1989) 273.}
\REF\EPii{
	M.~Bando, T.~Kugo, N.~Maekawa and H.~Nakano
	\journal \PTP &90 (93) 405.
}
\REF\AC{
	T.~Appelquist and J.~Carazzone
	\journal \PR &D11 (75) 2856;
\nextline
	See also, K.~Symanzik
	\journal \CMP &34 (73) 7.
}
\REF\KY{
	Y.~Kazama and Y.-P.~Yao
	\journal \PRL &43 (79) 1562;
\nextline
	\andjournal \PR &D21 (80) 1116; \andvol &D21(80)1138;
	\andvol &D25 (82) 1605;
\nextline
	C.~K.~Lee   \journal \NP &B161 (79) 171;
\nextline
	T.~Hagiwara and N.~Nakazawa \journal \PR &D23 (81) 959.}
\REF\Wein{
	S.~Weinberg  \journal \PL &91B (80) 51;
\nextline
	B.~Ovrut and H.~Schnitzer \journal \PR &D21 (80) 3369;
	\andvol &D22 (80) 2518; \andjournal \NP &B179 (81) 381;
\nextline
	P.~Bin{\'e}truy and T.~Sch{\"u}cker
	\journal \NP &B178 (81) 293; 307;
\nextline
	H.~Georgi and S.~Dawson \journal \NP &B179 (81) 477.}
\REF\WITTEN{
	E.~Witten \journal \NP &B104 (76) 445;
\nextline
	Y.~Kazama and Y.-P.~Yao
	\journal \PR &D25 (82) 1605
	and references therein.}
\REF\GEORGI{
	H.~Georgi \journal \NP &B361 (91) 51.}
\REF\GP{H.~Georgi and S.~Politzer
	\journal \PR &D14 (76) 1829.}
\REF\TW{G.~'t Hooft \journal \NP &B61 (73) 455;
\nextline
	S.~Weinberg \journal \PR &D8 (73) 3497.}
\REF\MDRGE{
	M.~Bando, T.~Kugo, N.~Maekawa and H.~Nakano
	\journal \PR &D44 (91) 2957.}
\REF\WETT{
	C.~Wetterich
	\journal Z.~Phys. &C48 (90) 693;
	\andvol &C58 (93) 585.
}
\REF\EJ{
	M.~B.~Einhorn and D.~T.~Jones
	\journal \NP &B230[FS10] (84) 261.}
\REF\GRZ{
	G.~Gamberini, G.~Ridolfi and F.~Zwirner
	\journal \NP &B331 (90) 331.}
\REF\OYY{
	Y.~Okada, M.~Yamaguchi and T.~Yanagida
	\journal \PTP &85 (91) 1;
\nextline
	R.~Barbieri, M.~Frigeni and F.~Carvaglio
	\journal \PL &B258 (91);
\nextline
	H.~E.~Haber and R.~Hempfling
	\journal \PRL &66 (91) 1815.}
\REF\TV{
	G.~'t Hooft and M.~Veltman
	\journal \NP &B153 (79) 365.}
%%%%%%%%%%%%%%%%%%%%%%%%%%%%%%%%%%%%%%%%%%%%%%%%%%%%%%%%%%%%%%%%%
%%%%%%%%%%%%%%%%%%%%%%%%%%%%%%%%%%%%%%%%%%%%%%%%%%%%%%%%%%%%%%%%%

\tolerance=1000
%\nopubblock
\KUNS={1216}
\HETH={93/09}

\titlepage

\title{
Improving the Effective Potential,
Multi-Mass Problem
\break and Modified Mass-Dependent Scheme}

\author{ Hiroaki~NAKANO and Yuhsuke~YOSHIDA }

\Kyoto

\abstract{
We present a new procedure for improving the effective potential
by using renormalization group equation (RGE) in the presence of
several mass scales.
We propose a modification of the mass-dependent (MD) renormalization
scheme, $\MDB$ scheme, so that the scalar mass parameter runs at most
logarithmically on the one hand and the decoupling of heavy particles
is naturally incorporated in the RGE's on the other.
Thanks to these properties, the procedure in $\MDB$ scheme turns out
to be very simple compared with the regionwise procedure in $\MS$
scheme proposed previously.
The relation with other schemes is also discussed both analytically
and numerically.
}

\endpage

\chapter{Introduction}

Recently,
there have been the renewed interests on
how to sum up large logarithms in the effective potential,
to investigate the standard model and beyond.
Basically,
large logarithms like $\ln(M/\mu )$,
which makes the perturbation expansion unreliable,
appear when one deals with a system possessing
large mass scale $M$ compared with the scale $\mu $
at which one discuss the physics.
In this situation
one considers resumming the perturbation series
by using the renormalization group equation\rlap.\refmark{\GQW}
When one concerns with the functional form of the effective potential,
one considers its renormalization-group (RG) improvement.
This is well-known since the work by Coleman and Weinberg\refmark{\CW}
for the massless $\lambda \phi ^4$ theory,
although the complete description
even for the massive $\lambda \phi ^4$ theory has been given
only recently\rlap.\refmark{\KASTENING,\EPi,\FJSE}

In many realistic applications,
one often has to deal with an additional mass scale $m$
with the hierarchy $\mu \ll m\ll M$.
In the supersymmetric standard model, for instance,
one can regard $\mu $, $m$ and $M$ as
the weak scale, supersymmetry breaking scale and unification scale,
respectively.
When we discuss such a system,
we face the problem of multi-mass-scales\rlap:\refmark{\SHER}
there appear several types of logarithms,
$\ln(M/\mu )$ and $\ln(m/\mu )$,
while we are able to sum up
just a single logarithm by using the RGE.

In Ref.~[\EPii],
one way to improve the effective potential
in the presence of multi-mass-scales
was described in $\MS$ renormalization scheme.
The point was to make use of the decoupling
theorem\refmark{\AC-\GEORGI} and to divide the energy region
(region of field space) so that in each region,
there remains essentially a single log factor.
Although there is nothing wrong in principle,
such regionwise procedure may be cumbersome in practice.
So it is desirable to have an alternative way
to handle multi-mass-scale systems.

In this paper,
we propose a simple modification of
the conventional mass-dependent (MD) renormalization scheme,
which we call {\it modified MD scheme} ($\MDB$ scheme),
and apply it to improving the effective potential
in the presence of several mass scales.
Basically in MD scheme, the RG coefficient functions ($\beta $ and
$\gamma $ functions) depend on mass parameters
and hence the decoupling of heavy particles is taken into account
in the form of RG runnings\rlap.\refmark{\GP}
In addition, the proposed $\MDB$ scheme has a property that
mass parameters run at most logarithmically
while keeping the `automatic' decoupling in the RGE's; namely,
it enjoys simultaneously that
\item{(i)}~
the quadratic running of the scalar mass parameters is absent,
\item{(ii)}~
the decoupling effects of heavy particles are naturally built in.

\noindent
Based on these properties $(i)$ and $(ii)$,
we show, by adopting a simple model with two mass-scales, that
{\it the same condition as in the single-mass-scale
case}\refmark{\EPi} {\it is enough to achieve the RG improvement of
the effective potential over the whole region of field space}.

We should remark that
the property $(i)$ is crucial to prove the above statement.
Generally in MD scheme,
there appear non-logarithmic and power-like corrections
proportional to $\m$,
which are potentially large in the high-energy region.
Such non-logarithmic corrections cause trouble
in summing up the leading logs.
We modify the renormalization scheme
in order to cure this point.

The existence of non-logarithmic corrections is related to
the scheme dependence of the RG improved potential.
[To examine this point is another motivation of the present work.]
Note that
it is not trivial at all that
the RG improved potentials in $\MS$ and $\MDB$ schemes
coincide with each other.
Of course, the {\it full} effective potential
is independent of the renormalization scheme:
the effective potentials in various schemes are related
with each other simply by changes of variables.
The effective potential correctly calculated
up to a certain loop order is also scheme independent
since the loop expansion has a scheme-independent parameter, Planck
constant $\hbar$.
In general, however,
once one makes an approximation to the full theory,
it is quite possible that
the results are different scheme by scheme;
some schemes give better approximations
than the others.

In our case,
we approximate the full effective potential
by resumming `logarithmic' parts of the perturbation series
so that it satisfies the RGE.
Then the scheme independence becomes nontrivial:
the RGE relates `log factors' at different loop orders,
but the `logarithmic' structure will differ scheme by scheme.
Moreover,
there may appear non-logarithmic corrections
as mentioned above.
This is why
there is no {\it a priori} relation
between the RG improved potentials in various schemes.
Do they give the same approximations?
This is the problem of the scheme dependence.
Our result will support to some extent the naive expectation that
it {\it is} scheme independent.

This paper is organized as follows.
In the next section,
we briefly review the basic ingredients
for improving the effective potential by the RGE.
In sections~3 and 4,
we define the $\MDB$ renormalization scheme
and discuss its basic features.
We show that
the RGE's in $\MDB$ scheme inherit the nice property $(i)$
as in mass-independent (MI) scheme
	\refmark{\TW}
as well as the property $(ii)$ as in MD one.
The absence of the quadratic running is proved
directly from the renormalization conditions
and the automatic decoupling is established
by utilizing the decoupling theorem.
The detailed study on the structure of the effective potential in
$\MDB$ scheme is given in section~5.
We first define the leading log series expansion
in $\MDB$ scheme
and describe how to sum up the leading log.
It will be shown,
by examining the high- and low-energy regions separately,
that we can correctly sum up all log factors over the whole region
and that non-logarithmic corrections are in fact small.
After establishing the procedure in $\MDB$ scheme,
we compare the leading log potential in $\MDB$ scheme
with those in other schemes such as $\MS$ and MD ones
in section~6.
By numerically solving the RGE's,
a good coincidence will be found.
A final section is devoted to conclusions
and further comments.
Some one-loop results can be found
in appendices.

\chapter{Improving the Effective Potential in $\MS$ scheme\break
and Problems of Multi-Mass-Scales}

In order to explain the basic ingredients needed later, let us first
make a review of the procedure\refmark{\EPi,\EPii} for improving the
effective potential by using the RGE in $\MS$ scheme.
We also describe why the problem of multi-mass-scales arises
in the context of RG improvement of the effective potential.

Following Ref.~[\EPii],
let us consider the Yukawa model
$$
{\cal L} = \1{2}(\p \phi )^2 - \1{2}m^2 \phi ^2
	-\1{4!}\lambda \phi ^4
	+ \widebar\psi (i \rlap/\p -g\phi )\psi
	- \omega
\eqn\eqMODELi$$
where $\phi $ is a massive real scalar field
and $\psi =(\psi _1, \cdots ,\psi _N)^T$ are massless Dirac fields.
We take the Dirac field to be $N$-component
in order to indicate
which correction comes from fermion loop.
For simplicity, we impose `chiral-parity' invariance,
$\phi  \rightarrow  -\phi , \ \psi  \rightarrow  \gamma _5 \psi $,
to forbid the bare mass of fermion.
The last term $\omega $($=hm^4$ in the notation in Ref.~[\EPi])
is a vacuum-energy term,
which is usually omitted
but plays an important role
	\refmark{\EPi} in $\MS$ scheme.
In this paper, we assume that both coupling constants $g^2/(16\pi ^2)$
and $\lambda /(16\pi ^2)$ are small and of the same order.

In order to compute the effective potential $V(\rp)$
for the scalar field VEV, $\rp =\VEV{\phi }$,
we make a field shift $\phi  \rightarrow  \phi + \rp$
in \eqMODELi, and obtain
$$\eqalign{
{\cal L}= 	&\1{2}(\p \phi )^2 - \1{2}\MB \phi ^2
	 - \1{3!} \lambda \rp \phi ^3 - \1{4!} \lambda \phi ^4 \cr
	&+ {\widebar\psi }{\big(}
		i \rlap/\p - M_{{\rm {\sss F}}}- g\phi  {\big)}\psi
	 - V^{(0)}(\rp)
	 + [~\hbox{$\phi$-linear terms}~] \cr
}\eqn\eqMODELii$$
where the last term is the tree potential
$V^{(0)}(\rp) \equiv  \omega  + m^2{\rp}^2/2 + \lambda {\rp}^4/4!$
and we have introduced the masses
for the boson $\phi $ and fermion $\psi $, respectively,
$$
\MB \equiv  m^2 + \2{\lambda }{\rp}^2 \ , \qquad
M_{{\rm {\sss F}}} \equiv  g \rp
\eqn\eqMASS$$
in the presence of the scalar background $\rp$.
When $\rp$ is small,
the field $\phi $ may be regarded as heavy field
and $\psi $ as light field.

The starting point is that
the effective potential is independent of the renormalization point
$\mu$ and thus satisfies the RGE
$$\eqalignno{
0 &= \mu {d \over d\mu }V
   = {\cal D} V(\rp,\lambda ,g^2,m^2,\omega  \,;\mu ) \ ,
&\eqname\eqPOTRGE \cr
\noalign{\vskip 0.5cm}
{\cal D} &\equiv  \mu \der\mu
	+ \beta _\lambda  \der{   \lambda   } + \beta _g  \der{  g^2  }
	- \gamma _m  \der{\ln m^2} - \gamma _\phi  \der{\ln \rp}
	- \gamma _\omega  \,m^4\der{\omega }
&\eqname\eqRGOPERATOR \cr
}$$
supplemented with the RGE's for parameters,
such as
$$
\mu  {d \over d\mu } g^2 =   \beta _g     \ , \quad
\mu  {d \over d\mu } m^2 = - \gamma _m m^2 \ , \quad
\mu  {d \over d\mu } \omega   = -\gamma _{\omega }m^4 \ .
\eqn\eqRUN$$
One can immediately write down the general solution
to \eqPOTRGE\ as
$$V(\rp,\lambda ,g^2,m^2,\omega  \, ;\mu ) =
V{\big(}{\widebar \rp}  (t),{\widebar \lambda }(t),{\widebar  g }^2(t),
	{\widebar  m }^2(t),{\widebar \omega }(t) \, ;
	e^t \! \mu  {\big)}
\eqn\eqFORMALSOLUTION$$
where the barred quantities ${\widebar \rp}(t)$, \etc,
denote the solutions of running equations
with a running distance $t$ from the initial values $\rp$, \etc,
at the renormalization point $\mu $.
[Here we are regarding the RGE's
as differential equation
with respect to an independent `time' $t$, not $\mu $.]

The RG improvement of the effective potential
consists in solving the RGE \eqPOTRGE.
The RGE \eqPOTRGE\ by itself, however,
does not determine the RG improved effective potential
since it is the first order {\it homogeneous} differential equation.
We should impose the suitable boundary condition on the functional
form of $V$ at a certain `time' $t$.
We call the boundary `value' of the potential {\it boundary function}.
The RG improved potential is fixed
by requiring that the R.H.S. of Eq.~\eqFORMALSOLUTION\ coincide
at a certain `time' $t$ with the boundary function.
[The RGE \eqPOTRGE\ guarantees that
we can make a convenient choice of $t$.]
It is the choice of boundary functions
that determines how well
the obtained potential approximates the exact one.

How can we find a suitable boundary function ?

Let us work in $\MS$ scheme for a moment and examine
the detailed structure of loop corrections to the effective potential.
[The following arguments are valid in any mass-independent (MI)
schemes since the structure of loop corrections does not change.]
The simplest way\refmark{\EPi} to see this is
to rewrite our Lagrangian \eqMODELii,
by rescaling the fields by a factor $g$
as $\Phi =g\phi $ and $\Psi =g\psi $, into
$$\eqalign{
{\cal L}= \1{g^2}
	&\bigg[
	   \1{2} (\p \Phi )^2 - \1{2} \MB \Phi ^2
	 + {\widebar \Psi }( i \rlap/\p - M_{{\rm {\sss F}}} )\Psi  \cr
	&- \1{3!}{\big(}{\lambda \over g^2}
		M_{{\rm {\sss F}}} {\big)} \Phi ^3
	 - \1{4!}{\big(} {\lambda \over g^2} {\big)} \Phi ^4
	 - {\widebar \Psi } \Phi  \Psi
	\bigg] - V^{(0)}(\rp) \cr
}\eqn\eqMODELiii$$
and to regard $\Phi $ and $\Psi $ as our basic quantum fields.
In this form, the parameter $g^2$ is
an overall factor in front of the action
just like Planck constant $\hbar $.
So, $L$-loop contribution ($L\geq 1$) to the effective potential
clearly takes the form:
$$
V^{(L)} = g^2{\rp}^4 \cdot {\big(} \gpai {\big)}^L \times
	{\Big[}~\hbox{function in } \, \ln{\MF\over \m},
		\ \ln{\MB\over \m}, \
		{\MF\over \MB}, \ {\lambda \over g^2}~{\Big]} \ .
\eqn\eqVLMS$$
In $\MS$ scheme, we have two types of logarithms
$\ln(\MF/\mu ^2)$ and $\ln(\MB/\mu ^2)$
in our two mass-scale system,
both of which can become large.
Since we know the logarithms appear
at most to $L$-th power at $L$-loop level,
we can rewrite Eq.~\eqVLMS\ into
$$V^{(L)} = { \MFF \over g^2}\, \sum_{i,j\geq 0}^{i+j\leq L}
	    {\big(}\gpai{\big)}^{L-(i+j)} \,
	    v^{(L)}_{i,j}(x,y)\, \sF^i \sB^j
\eqn\eqVLiiMS$$
by introducing the variables
$$\eqalignno{
\sF &\equiv  \gpai \ln{\MF \over \m}~, \quad
\sB  \equiv  \gpai \ln{\MB \over \m}~,
&\eqname\eqVARiMS \cr
 x  &\equiv  {\MF \over \MB}~, \qquad\qquad\quad
 y   \equiv  {\lambda   \over g^2} \ .
&\eqname\eqVARiiMS \cr
}$$
Although we are assuming that
the coupling constant $g^2/(16\pi ^2)$ is small,
we should regard the `Kastening variables' $\sF$ and $\sB$ as
$O(1)$ since the logarithms may be large.
Other variables $x$, $y$ and hence
the coefficient functions $v^{(L)}_{i,j}(x,y)$ are also $O(1)$.
Then we sum up $V^{(L)}$ with respect to $L$
and further rewrite it into the summation over $\l\equiv L-(i+j)$ as
$$\eqalignno{
V = \sum_{L=0}^{\infty } V^{(L)}
  &= \omega  + { \MFF \over g^2}\,
	\sum_{\l=0}^{\infty } \ {\big(}\gpai{\big)}^{\l} f_{\l}\ ,
&\eqname\eqLLSEIRESMS \cr
f_{\l}(\sF, \sB \, ; x,y)
  &\equiv  \sum_{i,j\geq 0}^{\infty }
	v^{(\l+i+j)}_{i,j}(x,y)\, \sF^i \sB^j
&\eqname\eqLLFUNCTIONMS \cr
}$$
where we have included the tree part,
$f_0=x^{-1}/2 - 5y/24$, into the summation.
This form of the expansion of the effective potential,
first introduced by Kastening\refmark{\KASTENING} for the single
mass-scale case, is called {\it leading log series expansion}.
When expressed in terms of the variables \eqVARiMS\ and \eqVARiiMS,
it is the power series expansion
in the small coupling constant $g^2/(16\pi ^2)$.
The coefficients $f_0, f_1, \cdots , f_{\l}, \cdots $ correspond to
the leading, next-to-leading,
$\cdots $, $\l$-th-to-leading, $\cdots $ log terms, respectively.
Of course,
it does not matter whether one uses $\lambda $ instead of $g^2$
as the expansion parameter.

Now let us return to the question of
how to specify the boundary function.
The summation in \eqLLFUNCTIONMS\
for the $\l$-th-to-leading log term $f_{\l}$ involves
the quantities at $L=\l,\l+1,\cdots $ loop level.
If one could set $\sF = \sB = 0$ in \eqLLFUNCTIONMS,
then only the first term with $i=j=0$ would survive
and the summation would ternimate at finite loop order, $L=\l$,
$$f_{\l}(0, 0 \, ; x,y)
  =  v^{(L=\l)}_{0,0}(x,y) \ .
\eqn\eqBOUNDARYFUNCTIONMS$$
Since $v^{(L)}_{0,0}$ can be obtained
by computing $L$-loops,
this would imply that
one could use the $\l$-loop potential,
$V_{\l}=V^{(0)} + \cdots + V^{(\l)}$,
evaluated at $\sF=\sB=0$
as the boundary function for the $\l$-th-to-leading log potential.
In other words, if one could find a `time' $t_0$ such that
$$
\sFB (t_0) = \sBB (t_0) = 0 \ ,
\eqn\eqCONDITIONMSstrong$$
then the desired $\l$-th-to-leading log potential would be given
\foot{
As was {\it proved} in Ref.~[\EPi],
one should use the RGE's at $(\l+1)$-loop order.
Note also that, strictly speaking,
the error in this equation is $O(({\wb g}^2/16\pi ^2)^{\l+1})$,
not $O((g^2/16\pi ^2)^{\l+1})$,
but the difference will be small
unless some coupling blows up
(where our approximation itself does not make sense).
}
by requiring that the R.H.S. of Eq.~\eqFORMALSOLUTION\ be
the boundary function $V_{L=\l}|_{\sF=\sB=0}$:
$$\eqalign{
V (\rp,\lambda ,g^2,m^2,\omega  \, ;\mu )
=V_{L=\l}&
 {\big(}{\widebar \rp}(t),{\widebar \lambda }(t),{\widebar  g }^2(t),
 {\widebar  m }^2(t),{\widebar  \omega }(t)\, ; e^t \! \mu  {\big)}
 {\bigg|}_{\scriptstyle \sFB (t)=0 \atop \scriptstyle \sBB (t)=0} \cr
+&\,{ {\wb M}_{{\rm {\sss F}}}^4(t_0) \over {\wb g}^2(t_0) }\,
 \times \,O{\bigg(}{\Big(}{{\wb g}^2(t_0) \over 16\pi ^2} {\Big)}^{\l+1}
	   {\bigg)} \ . \cr
}\eqn\eqLLPOTENTIALMS$$

Actually,
the condition \eqCONDITIONMSstrong\ is sufficient,
but not necessary one.
For our purpose, it would be enough to find a `time' $t_0$
at which the logarithm factors $\sFB(t_0)$ and $\sBB(t_0)$ are
of $O({\widebar g}^2/16\pi ^2)$, instead of zero,
$$\sFB (t_0) \simeq \sBB (t_0)
= O{\Big(} { {\widebar g}^2(t_0) \over 16\pi ^2 } {\Big)} \ .
\eqn\eqCONDITIONMSweak$$
As can be seen from Eq.~\eqVLiiMS, under this condition,
these log factors contained at $L$-loop level reduce to
precisely $L$-th-to-leading log order quantities.
So, to obtain the boundary function
for $\l$-th-to-leading log potential,
it would be necessary and sufficient
	\refmark{\EPii}
to retain these log factors up to $\l$-loop,
just as in Eq.~\eqLLPOTENTIALMS.

Unfortunately, such condition \eqCONDITIONMSstrong,
or even weaker one \eqCONDITIONMSweak,
can not always be satisfied simultaneously
since the difference $\sB - \sF$ becomes $O(1)$
when $g^2{\rp}^2 \ll  m^2$.
The RGE \eqPOTRGE\ enables us to set just a single variable
to desired value, but not several variables.
So one can not find a solution
to \eqCONDITIONMSstrong\ or \eqCONDITIONMSweak\
and is left with the infinite summation in \eqLLFUNCTIONMS.
This is the problem of `multi-mass-scales'
in the context of the RG improvement of the effective potential.

Now, we examine whether MD scheme provides us with a solution to this
problem.
Even in MD scheme, one will have a similar structure of the leading
log series expansion as in Eq.~\eqLLSEIRESMS.
However, there explicitly appears the renormalization point $\mu ^2$
in the effective potential, other than $\ln\mu $.
Such explicit $\mu $ dependence is closely related to
the existence of the quadratic running of scalar mass
and makes the perturbation theory unreliable.
In particular, in the context of the leading log series expansion,
it may make the coefficient functions $v^{(L)}$ arbitrarily large.
In fact, as we shall show in the following sections,
we remedy this point by modifying the renormalization conditions in MD
scheme.

\chapter{Modified Mass-Dependent Scheme}

In this section,
we give a definition of the modified MD scheme ($\MDB$ scheme).
For the theory \eqMODELi,
we define the $\MDB$ renormalization scheme
by the following renormalization conditions.
For the scalar two-point vertex $\Gamma ^{(2)}_{\phi }$,
we impose
$$\eqalignno{
&\Gamma _\phi ^{(2)}
	{\Big\vert}_{\scriptstyle p^2=0 \atop \scriptstyle m^2=0}
	\equiv  \lim_{m^2\rightarrow 0}
	   {\Big(} \Gamma _\phi ^{(2)}{\Big\vert}_{p^2=0} {\Big)}
	= 0 \ ,
&\eqname\eqRCi \cr
\noalign{\vskip 0.3cm}
&{\p \over \p m^2}\Gamma _\phi ^{(2)}{\bigg\vert}_{p^2=-\mu ^2}
	= - 1 \ ,
&\eqname\eqRCiim \cr
\noalign{\vskip 0.3cm}
&{\p \over \p p^2}~\Gamma _\phi ^{(2)}{\bigg\vert}_{p^2=-\mu ^2}
	= 1 \ .
&\eqname\eqRCiip \cr
}$$
The fermion two-point vertex takes the form
$\Gamma ^{(2)}_{\psi }=A(p^2)\,\rlap/p$
due to the `chiral-parity' symmetry, for which we require
$$
A{\Big\vert}_{p^2=-\m} = 1 \ .
\eqn\eqRCiii$$
Finally,
for the Yukawa vertex $\Gamma _g^{(3)}$
and the scalar four-point vertex $\Gamma ^{(4)}_{\phi }$,
we impose
$$
\Gamma _{g}^{(3)}{\big(}p, -p \,;0{\big)}
{\Big\vert}_{p^2=-\m} = -g \ , \quad
\Gamma _{\phi }^{(4)}
{\Big\vert}_{p_ip_j=-\m\delta _{ij} + \1{3}\m(1-\delta _{ij})}
	= -\lambda
\eqn\eqRCiv$$
where we set the boson-external momentum equal to zero
in $\Gamma ^{(3)}_g$.

To be precise,
these renormalization conditions \eqRCi-\eqRCiv\
should be supplemented with that for the zero-point vertex
$\Gamma ^{(0)}$
$$\Gamma ^{(0)}{\Big\vert}_{m^2=0}
= {\p \over \p m^2}\Gamma ^{(0)}{\bigg\vert}_{m^2=0} = 0 \ , \quad
  \1{2} \Big( {\p \over \p m^2} \Big)^2 \Gamma ^{(0)} = -h \ .
\eqn\eqRCv$$
Clearly,
the vacuum-energy term $\omega =hm^4$,
which played an important role in $\MS$ scheme,
is completely independent of the renormalization point
and is irrelevant for later discussions.
[Instead, one can simply impose $\Gamma ^{(0)} = -\omega $.]

The new set of the renormalization conditions \eqRCi-\eqRCiv\ is
a modified version of MD ones.
The modifications are made in the conditions
on the scalar two-point vertex, Eqs.~\eqRCi\ and \eqRCiim,
which take the place of a single condition
$$
\Gamma _\phi ^{(2)}{\Big\vert}_{p^2=-\m} = - \m -m^2
\eqn\eqRCMD$$
in the conventional MD scheme.
With this modification,
$\MDB$ scheme enjoys the properties announced in the introduction;
to be precise,
\item{(i)}~
In the high-energy region $\m\gg m^2$,
the RGE's in $\MDB$ scheme approach to those
in a certain mass-independent (MI) scheme.
In particular,
the mass parameter runs at most logarithmically;
\item{(ii)}~
In the low-energy region $\m\ll m^2$,
the decoupling effects are automatically taken into account
in the RGE's and the vertex functions.

\noindent
As we shall see,
it is crucial to separate
the condition \eqRCMD\ into Eqs.~\eqRCi\ and \eqRCiim\ in order to
realize the property $(i)$, which will play important roles in
section~5.

Before showing the properties $(i)$ and $(ii)$,
let us take a close look at the renormalization conditions
\eqRCi-\eqRCiv.
First,
they differ from those in MI scheme\rlap.\refmark{\TW}
In MI scheme, one treats one parameter family of theories
with different values of mass
and renormalizes them at certain value like $m^2=0$, $m^2=\m$;
one imposes, for instance for $\Gamma ^{(2)}_{\phi }$,
the condition \eqRCi\ and
$$
{\p \over \p m^2}\Gamma _\phi ^{(2)}
{\bigg\vert}_{\scriptstyle p^2=-\mu ^2 \atop \scriptstyle m^2=\,0~~}
 = - 1 \ , \qquad
{\p \over \p p^2}\Gamma _\phi ^{(2)}
{\bigg\vert}_{\scriptstyle p^2=-\mu ^2 \atop \scriptstyle m^2=\,0~~}
= 1  \ .
\eqn\eqRCMI$$
Clearly all the renormalization constants
are independent of the renormalized mass parameter $m^2$.
On the other hand,
in our $\MDB$ scheme,
we are still treating, in a sense, one parameter family
of theories with different values of mass
in order to impose the conditions \eqRCi\ and \eqRCiim.
With the renormalization conditions \eqRCiim-\eqRCiv, however,
the renormalization constants $Z_X$
for $X=\phi ,~\psi ,~m,~g$ and $\lambda $
generally depend on the ratio $\m/m^2$;
$$
Z_X =Z_X {\big(}\lambda ,g^2,\ln{m^2 \over \m} \,;\, \momu {\big)} \ .
\eqn\eqDEPENDENT$$
The RG coefficient functions ($\beta $ and $\gamma $ functions),
which are calculated from $Z$'s,
also depend on the mass parameter.

Secondly, the renormalization constants $Z_X$ are consistently
determined in $\MDB$ scheme.
A complication occurs only in the scalar two-point vertex while other
vertices can be treated in the same manner as in MD scheme.
Let us write the scalar two-point vertex as
$$
\Gamma _{\phi }^{(2)}(p,- p\,;m^2)
= Z_{\phi }p^2 - Z_mm^2 + \Pi (p^2\,;m^2)\ .
\eqn\eqTWOPOINT$$
As usual, the wave-function factor $Z_{\phi }$ is determined
by the condition \eqRCiip:
$$
Z_{\phi } = 1 - {\p \over \p p^2} \Pi {\Big\vert}_{p^2=-\m} \ .
\eqn\eqZphi$$
As for $Z_m$,
the renormalization conditions \eqRCi\ and \eqRCiim\ yield,
respectively,
$$\eqalignno{
\lim_{m^2\rightarrow 0} m^2 Z_m &= \Pi {\big\vert}_{p^2=m^2=0} \ ,
&\eqname\eqZmzero \cr
	\Big( 1 + m^2 {\p \over \p m^2} \Big) Z_m
&= 1 +  \Big[ \big( 1 - p^2 {\p \over \p p^2} \big)
		\big( {\p \over \p m^2} \Pi  \big)
	\Big] {\bigg\vert}_{p^2=-\m}
&\eqname\eqZmone \cr
}$$
where
we have used Eq.~\eqZphi\ in deriving Eq.~\eqZmone.
Observe that
Eq.~\eqZmone, being a differential equation,
does not completely determine $Z_m$.
This is most evident by noting that
a piece $Z_m\sim \m/m^2$ drops from the L.H.S. of Eq.~\eqZmone.
What determines this piece is precisely
Eq.~\eqZmzero.
Thus we see that
the renormalization condition \eqRCi\ provides
the condition~\eqRCiim\ with a boundary condition
and that all the renormalization constants are
uniquely determined in $\MDB$ scheme.

Now, let us look at one-loop examples and confirm that the properties
$(i)$ and $(ii)$ actually hold.
[See appendices for more details.]
The $\beta $ and $\gamma $ functions are given by
$$\eqalign{
&16\pi ^2 \gamma _{\phi }= ~~2N g^2 \ , \qquad
 16\pi ^2 \gamma _{\psi }= \2{g^2}K_\psi  {\Big(} \muom {\Big)} \ ,\cr
&16\pi ^2 \gamma _m\! = - 4N g^2 \ , \qquad\!
 16\pi ^2 \beta _g   =   6  g^4 K_g {\Big(} \muom {\Big)}
		 + 4N g^4 \ , \cr
&16\pi ^2 \gamma _{\omega }= 0 \ , \qquad\qquad~\,
 16\pi ^2 \beta _\lambda
	=   3 \lambda ^2 K_\lambda {\Big(} \muom {\Big)}
	 + 8N g^2 (\lambda  - 6 g^2) \ . \cr
}\eqn\eqRGEoneMDB$$
The functions $K_X(z)$ ($X=\psi ,~g,~\lambda $) are defined by
$$\eqalign{
K_\psi (z) &= 1 +~{2 \over  z}\,
	 - ~{2 \over z}\,{\big(} 1 +~\1{ z}\,{\big)} \ln(z+1) \ , \cr
K_g (z) &= 1 + {2 \over 3z}
	 - {4 \over 3z}  {\big(} 1 + \1{2z}  {\big)} \ln(z+1) \ , \cr
K_\lambda (z) &= 1 - {3 \over 2z}
	 \1{ \sqrt{1+3/z} } \,
	 \ln { \sqrt{1+3/z}+1 \over \sqrt{1+3/z}-1 } \cr
}\eqn\eqKDEF$$
which are normalized to be $1$
in the high-energy limit $z~(=\m /m^2)\rightarrow \infty $ and,
remarkably, vanish in the low-energy limit $z\rightarrow 0$
[See Fig.~1.]:
$$
K_X(z)~\rightarrow ~
\cases{ 1 &as~~$\hbox{$z \rightarrow  \infty $}$ \cr
	0 &as~~$\hbox{$z \rightarrow  0 $}$ \cr} \ .
\eqn\eqLIM$$
Recall that the terms proportional to $N$ come solely from the
light-particle (fermion) loops.
Others come from the heavy-particle (boson) loops.
The latter terms are accompanied by the functions $K_X$,
which have the property \eqLIM.
This is nothing but the decoupling of heavy particle loops,
as claimed in $(ii)$.
\figmark{1}

In MI (or $\MS$) scheme,
the RGE's do not have such a property of the automatic decoupling.
Instead, one has to switch from the full theory
to the low-energy effective theory.
In MI scheme\rlap,
\foot{
Here we adopt the renormalization condition as in Eq.~\eqRCMI.
If we renormalize $\Gamma _{\phi }^{(2)}$ at $m^2=\mu ^2$,
instead of $m^2=0$,
then the $\gamma _m$ in such MI scheme coincides
with that in $\MS$ scheme:
$16\pi ^2 \gamma _m =-4Ng^2 - \lambda $.
But the difference is not so important here.
}
the RGE's for the full theory are
$$\eqalign{
&16\pi ^2 \gamma _{\phi }= ~~2N g^2 \ , \qquad
 16\pi ^2 \gamma _{\psi }=  \2{g^2} \ ,\cr
&16\pi ^2 \gamma _m\! = - 4N g^2 \ , \qquad\!
 16\pi ^2 \beta _g   =   6  g^4 + 4N g^4 \ , \cr
&16\pi ^2 \gamma _{\omega }=  -\1{2} \ , \qquad\qquad~\,
 16\pi ^2 \beta _\lambda   =
	3 \lambda ^2 + 8N g^2 (\lambda  - 6 g^2)\ .\cr
}\eqn\eqRGoneMSHE$$
In the low-energy effective theory, we keep only the terms
proportional to $N$ in Eqs.~\eqRGoneMSHE\ and have the RGE's
$$\eqalign{
&16\pi ^2 \gamma _{\phi }= ~~2N g^2 \ , \qquad
 16\pi ^2 \gamma _{\psi }= 0 \ ,\cr
&16\pi ^2 \gamma _m\! = - 4N g^2 \ , \qquad\!
 16\pi ^2 \beta _g   =   4N g^4 \ , \cr
&16\pi ^2 \gamma _{\omega }= 0 \ , \qquad\qquad~\,
 16\pi ^2 \beta _\lambda   =   8N g^2 (\lambda  - 6 g^2)\ .\cr
}\eqn\eqRGoneMSLE$$
By comparing Eqs.~\eqRGEoneMDB\
with Eqs.~\eqRGoneMSHE\ and \eqRGoneMSLE,
one clearly sees that the RGE's in $\MDB$ scheme interpolate
those in MI scheme for the high- and low-energy regions.

At one-loop order,
the $\gamma _m$ in $\MDB$ scheme is the same as in MI scheme:
$$
\mu {d \over d\mu }m^2 = {4Ng^2 \over 16\pi ^2}\,m^2 \ ,
\eqn\eqMASSRUNMDB$$
which means that the mass parameter in $\MDB$ scheme runs
logarithmically, as claimed in $(i)$. [See Fig.~2.]
This is the result of our modification of the renormalization
conditions.
\figmark{2}
This is in sharp contrast to the case of the conventional MD scheme.
Indeed, with the MD renormalization condition \eqRCMD,
a fermion one-loop contribution to $\Gamma _{\phi }^{(2)}$ produces
\refmark{\MDRGE,\WETT}
a piece proportional to $\m /m^2$ in $\gamma _m$;
$$
\mu {d \over d\mu }m^2 = {4Ng^2 \over 16\pi ^2}\,(m^2 + \m)\ .
\eqn\eqMASSRUNMD$$
One sees that the running of the mass parameter
is completely different from that in Eq.~\eqRGoneMSHE\
since the second term on the R.H.S. of Eq.~\eqMASSRUNMD\
dominates in the high-energy region
(while it approaches to that in Eq.~\eqRGoneMSLE\
in the low-energy region).
If such quadratic running is present,
the RGE's never interpolate
the MI ones in the high- and low-energy region.

\chapter{Logarithmic Running and Automatic Decoupling in RGE}

In the last section,
we illustrated the properties $(i)$ and $(ii)$
of $\MDB$ scheme by one-loop examples.
We now present the general argument to show that
these properties hold to any loop order.

Basically the property $(i)$ follows from the fact that
we introduce the $\mu$ dependence only through the {\it dimensionless}
combinations of vertex functions $\Gamma ^{(n)}$ ($n\not=0$),
such as $(\partial /\partial m^2 )\Gamma ^{(2)}$.
In other words,
we never introduce the $\mu $ dependence in
the renormalization condition \eqRCi\ on $\Gamma ^{(2)}_{\phi }$
which has dimension two.
Since the dependence on $\mu $ is introduced only through quantities
which are at most logarithmically divergent,
we do not meet the quadratic dependence on $\mu $.

This property $(i)$ can be confirmed directly as follows.
First, the condition \eqRCiip\ determines the momentum dependence of
the two-point vertex to be
$$
\Gamma ^{(2)}_{\phi } = p^2 - m^2 - \m~c{\big(} {m^2 \over \m} {\big)}
	+ {(p^2+\m)^2 \over \m}~
	  f{\big(} {-p^2 \over \m}, {m^2 \over \m} {\big)} \ .
\eqn\eqTWOgeneral$$
The condition \eqRCiim\ implies that
the unknown function $c$ is independent of $m^2$,
and the condition \eqRCi\ determines it to be $f(0,0)$.
Note that the function $f$ should be nonsingular in the limit
$m^2\rightarrow 0$ since $\Gamma ^{(2)}_{\phi }$
has a massless limit.
Then the renormalized scalar two-point vertex
in $\MDB$ scheme takes the form
$$
\Gamma ^{(2)}_{\phi } = p^2 - m^2
	+ \m~{\bigg[}~
		{\Big(} {p^2+\m \over \m} {\Big)}^2~
		f{\big(} {-p^2 \over \m}, {m^2 \over \m} {\big)}
	- f(0,0)~{\bigg]} \ .
\eqn\eqTWOMDB$$
Now we use the RGE for $\Gamma ^{(2)}_{\phi }$
$$\eqalignno{
0 &= {\Big(}~{\hat {\cal D}}
	- 2 \gamma _{\phi } - \gamma _m m^2 \der{m^2}~{\Big)}
	\Gamma ^{(2)}_{\phi } \ ,
&\eqname\eqTWORGE \cr
\noalign{\vskip 0.5cm}
{\hat {\cal D}} &\equiv  \mu \der\mu
	+ \beta _\lambda  \der{\lambda } + \beta _g  \der{g^2} \ .
&\eqname\eqRGOPii \cr
}$$
Inserting the general form \eqTWOMDB\ into Eq.~\eqTWORGE\ and
taking a limit $m^2\rightarrow 0$ after setting $p^2=0$,
we obtain that
$$
0 = \lim_{m^2\rightarrow 0}m^2\,\gamma _m~
	 {\bigg[}~1 - \m \der{m^2}f(0,{m^2 \over \m})~{\bigg]} \ .
\eqn\eqMASSLIMITi$$
Since the quantity in the square bracket
does not vanish (at least perturbatively),
Eq.~\eqMASSLIMITi\ implies that
$$
0 = \lim_{m^2\rightarrow 0} m^2 \gamma _m \ .
\eqn\eqMASSLIMITii$$
Since the quadratic running of the mass parameter
corresponds to the behavior $\gamma _m\sim \m/m^2$,
Eq.~\eqMASSLIMITii\ proves the absence of
the quadratic running in $\MDB$ scheme.

Next, we turn to the property $(ii)$.
Let us examine the relation between
the full theory in the low-energy region
and the low-energy effective theory.
Here the low-energy effective theory is obtained from the full theory
by regarding heavy fields as external fields
(instead of quantum fields), \ie,
by taking out heavy-field internal lines.
What we want to prove is that in $\MDB$ scheme,
the full theory in the low-energy region
will automatically go over into the low-energy effective theory.

In order to find such relation,
we make use of the decoupling theorem:\refmark{\AC,\KY,\Wein}
the contributions due to heavy particles,
aside from those which
are suppressed by the inverse power of the heavy mass,
can be renormalized
into the parameters of the low-energy effective theory.
Let $\Gamma ^{(n)}$ be $n$-point vertex in the full theory
and ${\wt \Gamma }^{(n)}$ the corresponding vertex
in the low-energy effective theory.
[We denote the quantities in the low-energy effective theory
by the tilde.]
Then, according to the decoupling theorem,
when all the external momenta $p_i$ as well as
the renormalization point $\mu $ are small
compared with the mass $m$, \ie, for $|p_ip_j|,~\m\ll m^2$,
$$
\Gamma ^{(n)}{\big(} p_i, g, \lambda , m^2\,;\, \mu  {\big)}
=Z_{\phi }^{\2{b}} Z_{\psi }^f \,
{\wt \Gamma }^{(n)}
{\big(} p_i, {\wt g}, {\wt \lambda }, {\wt m}^2\,;\, \mu  {\big)}
	+O{\Big(} {p_ip_j \over m^2}, {\m \over m^2} {\Big)}
\eqn\eqDECOUPLEi$$
where $b$ and $2f$ stands for the number of
external bosons and fermions, respectively: $n=b+2f$.
$\Gamma^{(n)}$ does not depend on ${\wt m}^2$ except for $n=2$.

Originally,
the low-energy effective theory is not completely fixed
by specifying the Lagrangian itself.
So we fix it by imposing the {\it same} $\MDB$
renormalization conditions as \eqRCi-\eqRCiv.
Then, let us look at the scalar two-point vertex
$$
\Gamma ^{(2)}_{\phi } (p, -p)
=Z_{\phi } \, {\wt \Gamma }^{(2)}_{\phi } (p, -p)
	+  O{\Big(} {-p^2 \over m^2}, \muom {\Big)}
	\times  {\big[} -p^2~\hbox{or}~\m \,{\big]}
\eqn\eqDECtwo$$
where we have retained a factor $-p^2~\hbox{or}~\m$.
[$m^2$ never appear here.]
By differentiating this equation with respect to $p^2$
and setting $p^2=-\m$,
the conditions \eqRCiip\
for $\Gamma _{\phi }^{(2)}$ and ${\wt \Gamma }_{\phi }^{(2)}$ lead to
$$
Z_{\phi }=1+O{\Big(} \muom {\Big)} \ .
\eqn\eqWAVEbose$$
Similarly,
one can use the conditions \eqRCiii\ to show that
$$
Z_{\psi }=1+O{\Big(} \muom {\Big)} \ .
\eqn\eqWAVEfermi$$
Thus, the relation \eqDECOUPLEi\
between $\Gamma ^{(n)}$ and ${\wt \Gamma }^{(n)}$ reduces simply to
$$
\Gamma ^{(n)}{\big(} p_i, g^2, \lambda , m^2\,;\, \mu  {\big)}
= {\wt \Gamma }^{(n)}
{\big(}\,p_i, {\wt g}^2, {\wt \lambda }, {\wt m}^2\,;\,\mu \,{\big)}
	+ O{\Big(} {p_ip_j \over m^2}, \muom \,{\Big)} \ .
\eqn\eqDECOUPLEii$$
As for the dimensionless couplings $g$ and $\lambda $,
we set $p^2=-\m$ in Eq.~\eqDECOUPLEii\ with $n=3,4$
and use the conditions \eqRCiv\
in the full and the low-energy effective theories to obtain
$$
g = {\wt g} + O{\Big(} \muom {\Big)} \ , \qquad
\lambda ={\wt \lambda } + O{\Big(} \muom {\Big)} \ .
\eqn\eqDECgl$$
In this way, the finite renormalizations are not necessary
also in the coupling constants.
It remains to show that
the same is true for the mass parameters:
$$
m^2 = {\wt m}^2 + O{\big(} \m {\big)}
    = {\wt m}^2 \times {\Big[}~1+O{\Big(} \muom {\Big)}~{\Big]} \ .
\eqn\eqDECm$$
We use the general form of
the scalar two-point vertex $\Gamma ^{(2)}_{\phi }$, \eqTWOMDB,
from which we have, by setting $p^2=-\m$,
$$
\Gamma ^{(2)}_{\phi }{\Big\vert}_{p^2=-\m}= -\m  - m^2 - \m f(0,0) \ .
\eqn\eqRCMDonTWOMDB$$
We also have the same expression for ${\wt \Gamma }^{(2)}_{\phi }$.
Inserting both the expressions into Eq.~\eqDECOUPLEii\
with $n=2$ and $p^2=-\m$,
we have
$$
- \m - m^2 - \m f(0,0)
= - \m - {\wt m}^2 - \m {\wt f}(0,0) + \m~O{\Big(} \muom {\Big)}
$$
which is nothing but the desired result \eqDECm.
Thus,
once the low-energy effective theory is renormalized
by the same $\MDB$ conditions,
we no longer need the finite renormalization
relating the full theory in the low-energy region to the low-energy
effective theory; all parameters in $\MDB$ scheme
automatically go over into those in the low-energy effective theory.

Since the parameters in high- and low-energy theories
are related in a way described above,
it is easy to see that
the RG coefficient functions in both theories
are the same modulo $O(\m/m^2)$ corrections:
for $X=g,\lambda $ and $Y=\phi ,\psi ,m^2$, we have
$$
\beta _X={\wt \beta }_X + O{\Big(} \muom {\Big)} \ , \qquad
\gamma _Y={\wt \gamma }_Y + O{\Big(} \muom {\Big)} \ .
\eqn\eqDECinRGE$$
This completes the proof of the property $(ii)$.

It is instructive here to see
how the $\MDB$ scheme modifies the conventional MD one.
Let us apply the same reasoning as above in the proof of $(i)$,
to the conventional MD scheme.
Again,
due to the condition \eqRCiip, the renormalized two-point vertex takes
the form \eqTWOgeneral.
Now, the renormalization condition \eqRCMD\ determines
the unknown function $c{\big(}g^2,\lambda ;\,m^2/\m{\big)}$ to be zero:
$$
\Gamma ^{(2)}_{\phi \,{\rm {\sss MD}}}
	= p^2 - m^2 + O{\big(} (p^2+\m)^2 {\big)} \ .
\eqn\eqTWOMD$$
Inserting this into the RGE \eqTWORGE\ and setting $p^2=-\m$,
we obtain
$$
\gamma _m = - 2{\Big(} 1 + {\m \over m^2} {\Big)}\gamma _{\phi } \ .
\eqn\eqRGmMD$$
This clearly shows that
the mass parameter in MD scheme runs
quadratically in the high-energy region
(as long as the wave-function renormalization
$\gamma _{\phi }$ does not vanish).

Let us recapitulate this in some different way.
In MD scheme,
the scalar two-point vertex \eqTWOMD\ satisfies all the $\MDB$
renormalization conditions except the condition \eqRCi,
and we meet the quadratic running of
the mass parameter $m^2_{{\rm {\sss MD}}}$.
Now, we finitely renormalize the two-point vertex
so that the condition \eqRCi\ is satisfied.
By requiring the equality between the expressions
\eqTWOMD\ and \eqTWOMDB\ at $p^2=-\m$, we have
$$
m^2_{{\rm {\sss MD}}} =
m^2_{{\wb {\rm {\sss MD}}}} + \m f(0,0) \ .
\eqn\eqRELATION$$
This shows that
the mass parameter $m^2_{{\wb {\rm {\sss MD}}}}$ is
just the logarithmic part of $m^2_{{\rm {\sss MD}}}$:
we have succeeded in separating the logarithmic and quadratic parts
in $m^2_{{\rm {\sss MD}}}$
by dividing the MD renormalization condition \eqRCMD\ into
the $\MDB$ ones \eqRCi\ and \eqRCiim.
This is the property $(i)$.

In this way,
we conclude that
the present $\MDB$ scheme simultaneously enjoys
the `automatic' decoupling of heavy particles
(as in the conventional MD scheme)
and logarithmic RG running (as in MI scheme).

\chapter{Improving the Effective Potential in $\MDB$ scheme}

We now turn to our main task of how to improve the effective potential
in the presence of several mass scales by using the RGE in $\MDB$
scheme.
Let us first describe the structure of
the effective potential for the system \eqMODELiii\ in $\MDB$ scheme.
This can be done
by applying almost the same reasoning
as reviewed in section~2 for $\MS$ case.
A difference arises from the finite part of counter-terms:
\foot{
Loop corrections themselves
depend on $m^2$ only through the combination $\MB=m^2+\lambda {\rp}^2/2$
since we evaluate the potential in the background $\rp=\VEV{\phi }$.
But since the renormalization constants are determined
in the symmetric phase $\rp=0$,
counter-terms produce the dependence
on the $\ln (m^2/\m)$ and $\m/m^2$.
}
there appear another type of log factor $\ln(m^2/\m)$
and non-logarithmic dependence on $\mu $.
So $L$-loop contribution
now takes the form
$$\eqalign{
V^{(L)}={\MFF \over g^2} {\big(}\gpai{\big)}^L
	&\Big[~\hbox{$L$-th order polynomial in}~
	 \ln{\MF\over \m},~\ln{\MB\over \m},~\ln{m^2\over \m} \cr
	&~~\hbox{whose coefficients depend on}~
	 {\MF\over \MB},~{\lambda \over g^2},~{\m \over m^2}~ \Big]
	\ . \cr
}\eqn\eqVLi$$
Introducing the variables
$$\eqalignno{
\sF  = \gpai \ln{\MF \over \m}~, \quad
\sB &= \gpai \ln{\MB \over \m}~, \quad
s_m \equiv  \gpai \ln{m^2 \over \m}~,
&\eqname\eqVARi \cr
 x   = {\MF \over \MB}~, \qquad\qquad\quad
 y  &= {\lambda   \over g^2}~, \qquad\qquad\quad~~
 z  \equiv  {\m  \over m^2} \ ,
&\eqname\eqVARii \cr
}$$
we rewrite Eq.~\eqVLi\ ($L\geq 1$) into
$$V^{(L)} = {\MFF \over g^2}\sum_{i,j,k\geq 0}^{i+j+k\leq L}
	    {\big(}\gpai{\big)}^{L-(i+j+k)} \,
	    v^{(L)}_{i,j,k}(x,y,z)\, \sF^i \sB^j s_m^k \ .
\eqn\eqVLii$$
Then we {\it define} the leading log series expansion
in $\MDB$ scheme by
$$\eqalignno{
V  = \sum_{L=0}^{\infty } V^{(L)}
  &= \omega  + { \MFF \over g^2}\, \sum_{\l=0}^{\infty }
	 {\big(} \gpai {\big)}^{\l} f_{\l}\ ,
&\eqname\eqLLSEIRESMDB \cr
f_{\l} (\sF, \sB , s_m \, ; x,y,z)
	&= \sum_{i,j,k\geq 0}
	v^{(\l+i+j+k)}_{i,j,k}(x,y,z)\, \sF^i \sB^j s_m^k
&\eqname\eqLLFUNCTIONMDB \cr
}$$
where we have again included the tree part into the summation
and we have defined the order $\l$ of this expansion
as $\l\equiv L-(i+j+k)$.

The leading log series expansion \eqLLSEIRESMDB\
is the power series expansion
in a small coupling constant $g^2/(16\pi ^2)$, as before.
Compared with \eqLLFUNCTIONMS\ in $\MS$ scheme, however,
the coefficient functions $v^{(L)}_{i,j,k}$
have the dependence on the new variable $z=\m /m^2$
which potentially makes $v^{(L)}_{i,j,k}$ large.
So we do not know, at this stage, whether or not
the expansion \eqLLSEIRESMDB\ is sensible one;
we do not know whether or not
terms in the $(\l+1)$-th-to-leading log order
are smaller than those in the $\l$-th-to-leading log order.
Furthermore,
we have a log factor $s_m\sim \ln(m^2/\m)$
in addition to the `original' ones \eqVARiMS, $\sF$ and $\sB$.
So it is not obvious how one can
sum up these logarithms simultaneously by using the RGE
which can eliminate just a single variable.

At first sight,
the situation in $\MDB$ scheme appears to be worse
than in $\MS$ scheme due to the new variables $s_m$ and $z$.
What save the day are the nice properties
established in the last section.
Based on the properties $(i)$ and $(ii)$
as well as the fact that
the model has the well-defined massless limit,
we claim that the {\it single} condition
$$\sFB (t)=0
\eqn\eqCONDITION$$
is enough to determine the correct boundary function
{\it over the whole region of field space}.
The potentially dangerous variable $z=\m/m^2$ are in fact harmless
and the remaining logs $\sBB$ and $\smB$ are automatically summed up;
otherwise they decouple.
\foot{
This is why we have used the log factor of the {\it lightest} particle
in the condition \eqCONDITION.
}
As a result, with $\l$-loop effective potential $V_{L=\l}$
and $(\l+1)$-loop RGE's in $\MDB$ scheme at hand,
the $\l$-th-to-leading log potential is given simply by
$$\eqalign{
V(\rp,\lambda ,g^2,m^2,\omega  \, ;\mu ) = V_{L=\l}
&{\big(}
 {\widebar \rp}  (t),{\widebar \lambda }(t),{\widebar  g }^2(t),
 {\widebar  m }^2(t),{\widebar \omega }(t) \, ; e^t \! \mu  {\big)}
 {\Big|}_{\scriptstyle \sFB (t)=0} \cr
+& \, {{\wb M}_{{\rm {\sss F}}}^4(t_0) \over {\wb g}^2(t_0) } \,
 \times \, O{\bigg(} {\Big(} { {\widebar g}^2(t_0) \over 16\pi ^2 }
	{\Big)}^{\l+1} {\bigg)} \ . \cr
}\eqn\eqLLPOTENTIALMDB$$

The proof of the statement proceeds in a regionwise manner;
we divide the field space into
the large $\rp$ (high-energy) region $\grp\gg m^2$,
the small $\rp$ (low-energy) region $\grp\ll m^2$
and the intermediate region $\grp\sim m^2$,
and prove that the statement holds region by region.
We should stress here that
the final answer \eqLLPOTENTIALMDB\ does not require
to divide the region of field space
unlike the procedure in $\MS$ scheme.

First, the proof is rather simple
in the intermediate region $g^2{\rp}^2 \sim  m^2$.
In this region,
$\sBB$ and $\smB$ are of $O({\wb g}^2/16\pi ^2)$ from Eq.~\eqCONDITION\
and the weak condition as in Eq.~\eqCONDITIONMSweak\ is satisfied.
So, $\sBB$ and $\smB$ are already summed together with $\sFB$
by the single condition \eqCONDITION\
and the correct choice of the boundary function
is the same as in the single mass-scale case,
\ie, $V_{L=\l}{\big\vert}_{\sF=0}$.

Concerning the asymptotic regions ($g^2{\rp}^2 \gg  m^2$ and
$g^2{\rp}^2 \ll  m^2$),
we make an observation needed for the proof.
The boundary function is defined by setting $\m=\MF~(=\grp)$
and depends on $m^2$ only through the ratio $m^2/\m$.
It follows that {\it as far as the boundary function is concerned},
taking the high-energy limit $\m=\grp\rightarrow \infty $
is equivalent to considering the massless limit $m^2\rightarrow 0$
while the low-energy limit $\m=\grp\rightarrow 0$ corresponds to
the limit $m^2\rightarrow \infty $.

It should be noticed that
the validity of this equivalence between the high-energy and massless
limits heavily depend on the property $(i)$.
If $m^2$ run quadratically, then the ratio $m^2/\m$ would approach
to a finite value in the high-energy limit,
and the equivalence would break down.

\section{High-energy region ($g^2{\rp}^2 \gg  m^2$)}

Now, let us begin with the large $\rp$ region, $g^2{\rp}^2 \gg  m^2$.
The second log factor $\sBB$ becomes
asymptotically equal to the first logarithm $\sFB$
$$
\sBB - \sFB = {{\wb g}^2 \over 16\pi ^2}
	{\Big[}\,\ln {{\wb \lambda } \over 2{\wb g}^2} + \ln {\big(}\,
	 1 + {2{\wb m}^2 \over {\wb \lambda }{\wb {\rp}}^2}\,{\big)}\,
	{\Big]}
	 = O {\Big(} {{\wb g}^2 \over 16\pi ^2} {\Big)}
\eqn\eqSSEQUAL$$
as is the case in $\MS$ scheme\rlap.\refmark{\EPii}
[Recall $y=\lambda /g^2=O(1)$.]
So, setting $\sFB$ equal to zero is equivalent
to setting $\sBB$ equal to zero,
modulo the quantities of $O({\widebar g}^2/16\pi ^2)$
which is of the higher order in the leading log series expansion.
Physically,
we can regard the massive particle (here $\phi $) as massless.
Thus the condition \eqCONDITION\ automatically sums up
the second log factor $\sB$ as well as the first one $\sF$.

Next, let us make sure that
the variable $z=\m/m^2$ does not make
the coefficient functions $v^{(L)}$ large in the high-energy limit
$z\rightarrow \infty $.
To this end,
it is enough to see that
there is no positive power term in $z$
when the potential is expanded asymptotically in $z^{-1}(\ll 1)$.
As noted above,
this limit is equivalent
to the massless limit $m^2\rightarrow 0$ in our boundary function
and we know that
there arises no singularity in the latter limit.
This establishes that
power-like pieces of $\m/m^2$ vanish in this region.

As for the third log factor $s_m$,
we show that it does not contribute to the boundary function.
Since we know that the massless limit is regular,
the dangerous variable $s_m$ in this limit disappears in our boundary
function like
$$
\momu \ln \momu \rightarrow  0\ .
$$
Thus, we no longer need to sum up this log factor
in the high-energy limit $g^2\rp^2 = \m \gg  m^2$.
This establishes our claim in the large $\rp$ region.

\section{Low-energy region ($g^2{\rp}^2\ll m^2$)}

Next, we turn to the small $\rp$ region.
Now, $\sFB=0$ no longer implies $\sBB=0$
since their difference $\sBB-\sFB$ becomes $O(1)$.
Furthermore, $\smB$ is also large.
So we can not sum up these two log factors, $\sBB$ and $\smB$,
simultaneously.
Instead,
we shall show that in the low-energy limit $g^2{\rp}^2 \rightarrow 0$,
the log factors $\sB$ and $s_m$ as well as $z=\m/m^2$ decouple
so that we no longer need to sum them up.

This can be seen by using the decoupling theorem.
As in Eqs.~\eqDECgl, \eqDECm\ and \eqDECinRGE,
all the parameters as well as $\beta $ and $\gamma $ functions
of the full theory approach, in the low-energy limit,
to those of the low-energy effective theory.
In particular,
$$
\sF = {\wt {\sF}}+O{\Big(} \muom {\Big)} \ .
\eqn\eqDECsF$$
This property of the automatic decoupling holds
also for the effective potential itself.
In particular,
the boundary function $V{\big\vert}_{\sF=0}$ satisfies
$$
V{\big\vert}_{\sF=0} =
{\wt V}{\big\vert}_{\stF=0}
+ \MFF \times  O{\Big(} {g^2{\rp}^2 \over m^2} {\Big)}\ .
\eqn\eqDECOUPLEPOT$$
Here ${\wt V}$ is the effective potential
in the low-energy effective theory
fixed by the same $\MDB$ renormalization conditions
and we have replaced $\sF=0$ with $\stF=0$
by using the Eq.~\eqDECsF.
[One-loop example can be found in appendix~A.]
{}From this expression,
we see that
there is no contribution from
potentially large variables $\sB$ and $s_m$
as well as negative power terms in $\m/m^2$.
The first term ${\wt V}$ does not contain
loop effects due to heavy particle by definition.
Furthermore,
the remaining terms are small by itself.
Thus,
we need not sum up $\sB$ and $s_m$.
This establishes the correctness of our boundary function
in the low-energy asymptotic region $g^2{\rp}^2=\m\ll m^2$.
Together with the automatic decoupling $(ii)$
in $\beta $ and $\gamma $ functions,
this completes the proof of our assertion in this region.

In this way,
we establish the procedure to improve the effective potential.
The final answer is given by Eq.~\eqLLPOTENTIALMDB.

\chapter{Comparison with Other Schemes: Numerical Study}

In this section,
we work with the leading-log order
and demonstrate the results of the RG improvement.
At the leading log order,
we use the tree potential $V^{(0)}$ as the boundary function,
and one-loop RGE's.

Note that
our procedure is such that,
at individual point $\rp$ in field space,
the value $V(\rp)$ of the improved potential is evaluated
by Eq.~\eqLLPOTENTIALMDB.
Actually,
this is not economical since
we solve the running equations at each $\rp$.
As explained in Ref.~[\EPi],
one can avoid this duplication
by finding the value of $\rp$
corresponding to the solution to $\sFB(t)=0$
for each value of the running distance $t$.
Namely, as we solve the running equations,
we simultaneously obtain the value of the effective potential at
$$
{\rp}^2 = e^{2t}\m\1{{\wb g}^2(t)}
	{\Big[} {{\wb \rp}(0) \over {\wb \rp}(t)}{\Big]}^2 \ .
\eqn\eqRUNTIMEi$$
The last factor ${\wb \rp}(0)/{\wb \rp}(t)$ does not depend
on the initial value ${\wb \rp}(0)$.
We set $\mu =m$ for later convenience.
Then, Eq.~\eqRUNTIMEi\ becomes
$$
\ln {\grp \over m^2} = 2t
	+ \ln \left[\,{{\wb g}^2(0)\over {\wb g}^2(t)}\,
		{{\wb \rp}^2(0)\over{\wb \rp}^2(t)}\,\right] \sim   2t
\eqn\eqRUNTIMEii$$
so that the region $\grp \ggll m^2$ corresponds to $t \ggll 0$.

We compare the improved potential
in $\MDB$ scheme with those in $\MS$ and MD schemes.
For this purpose,
we should match the renormalized parameters in different schemes
in order to guarantee that we are treating the same system.
For the leading log potential,
the parameter relations should be exact also in the leading log order.
By using the fact that the improved potential exactly satisfies the
RGE \eqPOTRGE\ with one-loop $\beta $ and $\gamma $ functions,
we match the parameters at $\mu =m$.
Then, the parameter relations reduce to the tree-level ones
since $s_m\sim \ln(m^2/\m)=0$.

We now show the result of our numerical calculations.
We present it
in a moderate case (a): $g^2=0.55$, $\lambda =2.3$ and
in extreme cases (b): $g^2=0.5 $, $\lambda =2.5$
in which ${\wb \lambda }$ blows up,
and $(c)$: $g^2=0.7$, $\lambda =2.0$
in which the vacuum becomes unstable.
We set $m=1$ (as mass unit), $N=1$
and the vacuum energy to $\omega =0$
so that $V(\rp=0)=0$.
Fig.~3 shows the asymptotic behavior of the leading log potentials
in $\MDB$, $\MS$ and MD schemes for the case (a).
[Other cases are similar, so omitted.]
The horizontal and vertical axes are
$\ln({\rp}^2/m^2)$ and $\ln|V/m^4|$, respectively.
We find that the scheme dependence is quite small.
\foot{
A rather good coincidence is found
between $\MS$ and MD schemes.
It is not clear to us that
this persists to higher leading log order
since the reasoning presented in section~5
will not apply to the conventional MD scheme.
}
This can be understood from the fact that
since we are using the tree potential as the boundary function,
the asymptotic behaviors in the large and small $\rp$ regions
are mainly determined
by the quartic coupling $\wb \lambda $ and mass $\wb m^2$, respectively.
\figmark{3}
The behaviors of ${\wb g}^2(t)$ and ${\wb\lambda }(t)$
in $\MDB$ scheme are shown in Fig.~4.
The horizontal axis is the same as in Fig.~3.
The scheme dependence is mild
even in the extreme case $(b)$ where
the quartic coupling $\wb \lambda $ blows up in the high-energy region
and the case (c) where
$\wb \lambda $ becomes negative and
the vacuum instability occurs.
\figmark{4}
\figmark{5}

In order to examine how much the RG improvement is obtained,
we evaluate the difference of the improved potential $V$
from the tree one $V^{(0)}$ normalized by $V^{(0)}$
$$
\chi (\rp)\equiv {V(\rp)-V^{(0)}(\rp) \over V^{(0)}(\rp)} \ .
\eqn\eqCHARACTER$$
Figs.~5 show the results of $\chi _{\MDS}$, $\chi _{\MSS}$ and
$\chi _{{\rm {\sss MD}}}$, respectively,
for each case of the parameter choices.
As expected,
the longer distance we run from $t=0$,
the larger improvement is obtained.
The large improvement $|\chi |\sim 1$ is obtained for $2t\sim 30$
since the variables $\sF$ and $\sB$ becomes $O(1)$
for our parameter choices.
In extreme cases,
even larger improvement is obtained,
but the leading log approximation itself breaks down for these cases.

Finally, we add a remark.
We can further improve the approximation
by requiring that the potential be correct
not only in the leading log order,
but also in one-loop level\rlap.\refmark{\EPi,\EPii}
For this combined approximation,
we should use the one-loop potential as the boundary function
while one-loop RGE's are enough.
Also,
the parameter should be matched at one-loop level
by performing the finite renormalization at $\mu =m$\rlap.
\foot{
By matching at $\mu =m$,
higher order terms are small in $\MS$ and $\MDB$ schemes.
This might not be the case in the conventional MD scheme.
}

\chapter{Conclusions and Discussions}

We have discussed the issues
concerning the RG improvement of the effective potential
in the presence of several mass scales.
Originally the coexistence of multi-mass-scales
causes trouble in determining the boundary function
needed for the general solution of the RGE.
By adopting a simple model possessing two mass scales,
we have seen in this paper that $\MDB$ scheme provides us with a
suitable choice of boundary functions without dividing the field
space:
the correct boundary function
for $\l$-th-to-leading log potential
is just $\l$-loop potential evaluated at $\sF=0$.

The $\MDB$ scheme is the new renormalization scheme
proposed in this paper.
Its crucial properties are
the automatic decoupling of heavy particles
and the absence of quadratic running of scalar mass.
These properties enable us to show that
the leading log series expansion is well-defined
even in the presence of non-logarithmic corrections.
Then,
the procedure can be stated
by a single condition over the entire region,
which is the same as in the single-mass-scale case.

We make a comment on other possible methods
to handle the multi-mass-scale systems.
First,
we already have the procedure in $\MS$ scheme\rlap.\refmark{\EPii}
Such regionwise procedure will be cumbersome
especially when there are many mass thresholds.
In $\MDB$ scheme,
calculations of the RGE's become harder,
but once they are calculated,
then various threshold effects are
automatically taken care of.
Our method will be more useful
when some intensive investigations will be needed,
such as scanning for large parameter space.
On the other hand,
Einhorn and Jones proposed
	\refmark{\EJ}
to introduce several renormalization points $\mu _i$
by which the multi-log factors are simultaneously summed up.
Their method is interesting,
but the RGE's become partial differential equations.
Our method described here
involves solving ordinary differential equations,
which will be much easier task.

Finally
we comment on possible applications of
the method in this paper.
Amongst, it is interesting to apply our procedure
to the analysis of the Higgs potential
in the supersymmetric standard model,
in which many mass scales are present.
When the supersymmetry breaking scale is rather high,
we expect large improvement to usual analyses
which make use of at most
one-loop potential\rlap.\refmark{\GRZ,\OYY}

\vskip 0.5cm

\ACK
The authors are greatly indebted to T.~Kugo
for many suggestions, patient discussions
and careful reading of our manuscript.
They are grateful to M.~Harada, Y.~Kikukawa and N.~Maekawa for comments.
They also thank to M.~Bando, H.~Hata and S.~Yahikozawa
for encouragements.

\Appendix{A}

We gather one-loop results
by the dimensional regularization.
We first determine the renormalization constants
$Z_X = 1 + \hbar Z_X^{(1)} + O({\hbar }^2)$
from $\MDB$ renormalization conditions \eqRCi-\eqRCiv.
We then calculate one-loop RGE's
and the one-loop contribution $V^{(1)}$ to the effective potential.
Our conventions	are:
$d=4-2\epsilon $, $1/\ebar \equiv 1/\epsilon -\gamma +\ln 4\pi $
and ${\rm Tr}\,{\bf 1}=4$.
A mass scale $\mu _0$ is introduced
so that the dimensionality becomes correct.
$\mu _0$ always appears as $\mu _0^{2\epsilon } \int d^dk/(2\pi )^di$
and is identified with the renormalization point $\mu $ in $\MS$ scheme.
Otherwise,
$\mu _0$ has nothing to do with $\mu $
which is introduced through the renormalization conditions.

As usual,
we renormalize the theory \eqMODELi\
in the symmetric phase, $\VEV{\phi }=0$.
First, the vacuum-energy at one-loop is
$$
\Gamma ^{(0)} = - \omega _{{\rm bare}}
	   + \1{4} \1{16\pi ^2} m^4
	    {\Big[}\,\1{\ebar} + \2{3}
		- \ln {m^2 \over \mu _0^2}\,{\Big]} \ .
\eqn\eqVACUUMENERGY$$
{}From $\omega _{{\rm bare}}=\omega +\hbar \omega ^{(1)}+ O({\hbar}^2)$,
the simplified condition $\Gamma ^{(0)}=-\omega $ leads to
$$
\omega ^{(1)} = \1{4} \1{16\pi ^2} m^4
	{\Big[}\,\1{\ebar} + \2{3} + \lnmzm\,{\Big]} \ .
\eqn\eqOMEGAone$$
Second, the boson self-energy,
$\Gamma _\phi ^{(2)}(p,-p) =
	Z_\phi \,p^2 - Z_m\,m^2 + \Pi (p^2)$, is given by
$$
\Pi (p^2)={2Ng^2 \over 16\pi ^2}\, p^2
	{\Big[}\,\1{\ebar} + 2 + \lnmzp\,{\Big]}
	+ {\lambda /2 \over 16\pi ^2}\, m^2
	{\Big[}\,\1{\ebar} + 1 + \lnmzm\,{\Big]} \ .
\eqn\eqSELFENERGYboson$$
The renormalization constant $Z_{\phi }^{(1)}$
is determined as usual by Eq.~\eqZphi\
while the $Z_m$ is determined
by solving the differential equation \eqZmone\
and imposing the boundary condition \eqZmzero\ to be
$$
Z_\phi ^{(1)} = - {2Ng^2 \over 16\pi ^2}
	{\Big[}\,\1{\ebar} + 1
	+ \ln {\mu _0^2 \over \m}\,{\Big]} \ , ~~~
Z_m^{(1)}  = {\lambda /2 \over 16\pi ^2}
	{\Big[}\,\1{\ebar} + 1 + \lnmzm\,{\Big]} \ .
\eqn\eqZpm$$
The fermion self-energy,
$\Gamma _{\psi }^{(2)}(p,-p)
= Z_\psi   \rlap/p - \Sigma (p)$, is similar:
$$
Z_\psi ^{(1)} =  - {g^2/2 \over 16\pi ^2}
		{\Big[}\,\1{\ebar} + \lnmzm
	      - I_\psi {\big(} \muom {\big)}\,{\Big]}
\eqn\eqZpsi$$
where $I_\psi (z) \equiv  (1 + 1/z)^2 \ln(z+1) - (1/z) - 2$.
The vertex correction to the Yukawa coupling is,
$\Gamma _g^{(3)}(p,-p\,;0) = -Z_g\,g + \Lambda _g(p,-p\,;0)$,
$$
\Lambda _g^{(3)}(p,-p\,;0)
	= g {g^2 \over 16\pi ^2}
		{\Big[} \1{\ebar} + \lnmzm
	      - I_g{\big(} {-p^2  \over m^2} {\big)} {\Big]}
\eqn\eqVERTEXyukawa$$
where $I_g (z) \equiv  (1 + 1/z) \ln(z+1)$.
{}From the condition \eqRCiv,
we obtain
$$
Z_g^{(1)}= {g^2 \over 16\pi ^2}
	{\Big[} \1{\ebar} + \lnmzm
	- I_g{\big(} { \m \over m^2} {\big)} {\Big]} \ .
\eqn\eqZg$$

The vertex correction to the quartic scalar coupling,
$\Gamma _{\phi }^{(4)} = -Z_{\lambda }\lambda  + \Lambda _{\lambda }$,
is a little bit complicated.
We separate the boson- and fermion-loop contribution,
$\Lambda _\lambda =\Lambda _{{\rm {\sss B}}}+\Lambda _{{\rm {\sss F}}}$.
Let $p_i$ be incoming external momenta ($i=1\,$-$\,4$)
and $s=(p_1+p_2)^2$, $t=(p_1+p_4)^2$ and $u=(p_1+p_3)^2$.
The boson-contribution consists of
the $s$-, $t$- and $u$-channel ones as
$\Lambda _{{\rm {\sss B}}}(p_i)
 = \Lambda _{{\rm {\sss B}}}(-s)
 + \Lambda _{{\rm {\sss B}}}(-t)
 + \Lambda _{{\rm {\sss B}}}(-u)$ where
$$\eqalignno{
\Lambda _{{\rm {\sss B}}}(-s)~~
&\equiv  {\lambda ^2/2 \over 16\pi ^2}
 {\Big[}\,\1{\ebar} + \lnmzm
	- I_\lambda {\Big(} {-s \over m^2} {\Big)}\,{\Big]} \ ,
&\eqname\eqQUARTICbosonDEF \cr
I_\lambda {\Big(} {-s \over m^2} {\Big)}
	&\equiv  \int\nolimits_0^1 d\alpha
	\ln {\Big[}\,1 + {-s \over m^2}\,\alpha (1-\alpha )\,{\Big]} \cr
	&= \ln {-s \over 4m^2}
	   + L{\Big(} \sqrt{1+{4m^2 \over -s}} + 1 {\Big)}
	   - L{\Big(} \sqrt{1+{4m^2 \over -s}} - 1 {\Big)}
&\eqname\eqINTdef \cr
}$$
where $L(\zeta ) \equiv \zeta \,(\ln \zeta  - 1)$.
The fermion-contribution is evaluated
in appendix~B:
$$\eqalignno{
\Lambda _{{\rm {\sss F}}}(p_i)
&= - {Ng^4 \over 16\pi ^2}~I_0(p_i)
 + [~\hbox{permutation in}~p_i~] \ ,
&\eqname\eqQUARTICfermion \cr
I_0(p_i)
&\equiv  \1{\ebar} + 2 + \ln{\mu _0^2 \over -(p_1+p_2)^2}
  + J{\big(}-p_1^2,\, -p_2^2~ ;\, -(p_1+p_2)^2{\big)} \cr
&~~~	- \1{\,4\,}~J{\big(}~
		p_1^2p_3^2\,,~p_2^2p_4^2~ ;~
		(p_1+p_2)^2(p_2+p_3)^2~{\big)}
&\eqname\eqQUARTICfermionDEF \cr
}$$
and the function $J(\xi ,\eta \,;\,\zeta )$
is defined in the appendix~B.
Both $\Lambda _{{\rm {\sss B}}}$ and $\Lambda _{{\rm {\sss F}}}$ are
completely symmetric in $p_i$ ($i=1$-$4$).
Then, we impose the condition \eqRCiv\
at the symmetric point, $p_i^2=-\m$ and $s=t=u=-(4/3)\m$, to obtain
$$
Z_{\lambda }^{(1)} \lambda
={3\lambda ^2/2 \over 16\pi ^2}
	{\Big[}\,\1{\ebar} + \lnmzm
	- I_{\lambda }{\Big(} {4\m \over 3m^2} {\Big)}\,{\Big]}
- {4!\,Ng^4 \over 16\pi ^2}
	{\Big[}\,\1{\ebar} + \ln {\mu _0^2 \over \m} + F~{\Big]}
\eqn\eqZlambda$$
with a constant $F=3.02198\cdots$.

Some remarks are in order.
First observe that
functions $I_X$ (contained in $Z_X^{(1)}$ for $X=\psi ,g,\lambda $)
behave in the high-energy (massless) limit
$\m/m^2\rightarrow \infty $ as
$$
I_X{\Big(} \muom {\Big)} \longrightarrow \ln \muom \ , \qquad
\d I_X{\Big(} \muom {\Big)} \longrightarrow 1 \ .
\eqn\eqHELIMofI$$
An important point is that
in the low-energy limit $\m/m^2\rightarrow 0$,
they all approach to constant values,
$I_{\psi }(0)=-1/2$, $I_g(0)=1$ and $I_{\lambda }(0)=0$, as
$$ I_X{\Big(} \muom {\Big)} \longrightarrow I_X(0) \ , \qquad
\d I_X{\Big(} \muom {\Big)} \longrightarrow 0      \ .
\eqn\eqLELIMofI$$
Note also that
we have to renormalize $\Gamma ^{(4)}_{\phi }$ at the symmetric point
in order that all the $s$-, $t$- and $u$-channel boson-loops
equally contribute to $I_{\lambda }$.

Now,
we turn to the RG coefficient functions and the effective potential.
The $\beta $ and $\gamma $ functions are obtained
by noting that the bare parameters are independent of $\mu $.
For instance,
from $g_{{\rm bare}}^2 = Z_g^2 Z_{\phi }^{-1} Z_{\psi }^{-2}g^2$,
$$\eqalign{
{\beta _g \over g^2}
	&= - \dmu \ln Z_g^2 Z_{\phi }^{-1} Z_{\psi }^{-2}
	 = 2 \d {\Big[}
	 - 2 Z_g^{(1)} + Z_{\phi }^{(1)} + 2 Z_{\psi }^{(1)}\,{\Big]}
	 + O{\big(} {\hbar}^2 {\big)} \cr
	&= {4Ng^2 \over 16\pi ^2}
	 + {2g^2 \over 16\pi ^2} \d {\Big[}\,
		2 I_g{\Big(} \muom {\Big)}
		+ I_{\psi }{\Big(} \muom {\Big)}\,{\Big]}
	 + O{\big(} {\hbar}^2 {\big)} \ . \cr
}$$
By looking at the asymptotic behavior \eqHELIMofI\ and \eqLELIMofI,
we introduce the `interpolating' functions \eqKDEF\
with the property \eqLIM\ by
$$\eqalign{
K_{\psi }(z)&\equiv z{d \over dz}I_{\psi }(z) \ , \qquad
K_{\lambda }(z) \equiv z{d \over dz}I_{\psi }{\big(}\3{4z}{\big)}\ ,\cr
K_g\, (z)&\equiv \3{1}\, z{d \over dz}
	{\Big[}\,2I_g(z) + I_{\psi }(z)\,{\Big]} \ . \cr
}\eqn\eqKbyI$$
With these definitions,
we finally obtain the RGE's \eqRGEoneMDB.

Next,
we discuss the effective potential $V(\rp)$ for $\rp =\VEV{\phi }$.
The one-loop contribution to the effective potential
takes the form $V^{(1)}= \VLOOP^{(1)} + \Delta V^{(1)}$ where
$$\eqalign{
\VLOOP^{(1)} =
	&- {4N \over 64\pi ^2} \MFF
	{\Big[} - \1{~\ebar~} - \2{3}
		+ \ln {\MF \over \mu _0^2}\,{\Big]} \cr
	&+ \1{64\pi ^2}    \MBF
	{\Big[} - \1{~\ebar~} - \2{3}
		+ \ln {\MB \over \mu _0^2}\,{\Big]}
}\eqn\eqVLOOP$$
while the contribution from counter-terms are,
from Eqs.~\eqOMEGAone, \eqZpm\ and \eqZlambda,
$$\eqalignno{
\Delta V^{(1)} \equiv
&~\omega ^{(1)} + Z_m^{(1)} \1{2} m^2{\rp}^2
	  + Z_{\lambda }^{(1)} \1{4!}\lambda  {\rp}^4 \cr
=&-{4N \over 64\pi ^2}\,\MFF\,
 {\Big[}\,\1{\ebar} + \ln{\mu _0^2 \over \m}\,{\Big]}
 -{4N \over 64\pi ^2}\,g^4{\rp}^4~F \cr
&+ \1{64\pi ^2}	   \, \MBF\,
 {\Big[}\,\1{\ebar} + \ln{\mu _0^2 \over \m} + \ln{\muom}\,{\Big]}
&\eqname\eqVCOUNTERMDB \cr
&+ \1{64\pi ^2}\,{\bigg[}\,
	\2{3}m^4 + m^2\lambda{\rp}^2-{\big(}\2{\lambda }{\rp}^2{\big)}^2
	I_{\lambda }{\Big(} {4\m \over 3m^2} {\Big)}\,
	{\bigg]} \ . \cr
}$$
Thus, the final form of $V^{(1)}$ in $\MDB$ scheme is
$$\eqalign{
V^{(1)}=
&-{4N \over 64\pi ^2}\,\MFF\,{\Big[}\,\ln{\MF \over \m}-\2{3}\,{\Big]}
 + \1{64\pi ^2}	   \,\MBF\,{\Big[}\,\ln{\MB \over \m}-\2{3}\,{\Big]} \cr
&+ \1{64\pi ^2}\,m^4	   {\Big[}\,\ln{\muom} + \2{3}\,{\Big]}
 + \1{64\pi ^2}\,m^2\lambda {\rp}^2{\Big[}\,\ln{\muom} + 1 \,{\Big]} \cr
&+ \1{64\pi ^2}\,{\big(}\2{\lambda }{\rp}^2{\big)}^2 {\Big[}\,
    \ln{\muom} - I_{\lambda }{\Big(}{4\m \over 3m^2}{\Big)} \,{\Big]}
 - {4N \over 64\pi ^2}\,g^4{\rp}^4~F \ .\cr
}\eqn\eqONEPOTMDB$$
Note that
the first two terms in Eq.~\eqONEPOTMDB\ take just the same form as
$V^{(1)}$ in $\MS$ scheme while the remaining terms are extra
contributions to it in $\MDB$ scheme.

{}From the expression \eqONEPOTMDB,
we can explicitly confirm what we have generally argued in section~5.
The expression \eqONEPOTMDB\ is suitable
in the large $\rp$ region.
In the low-energy region $\m\ll m^2$, we rewrite it,
by combining $\ln(\m/m^2)$ with $\ln(\MB/\m)$,
into the form
$$\eqalign{
V^{(1)}=
&-{4N \over 64\pi ^2}\,\MFF\,{\Big[}\,\ln{\MF \over \m}-\2{3}+F\,{\Big]}
 + \1{64\pi ^2}  \,\MBF\,{\Big[}\,\ln{\MB \over m^2}-\2{3}\,{\Big]}\cr
&+ \1{64\pi ^2}{\bigg[}\,
	\2{3}\,m^4 + m^2\lambda {\rp}^2
	- {\big(}\2{\lambda }{\rp}^2{\big)}^2
		\times  O  {\Big(} \muom {\Big)}\,{\bigg]} \cr
}\eqn\eqONEPOTMDBLEi$$
where we used $I_{\lambda }(0)=0$.
When the second term is expanded
in the small $\rp$ region ${\rp}^2\ll m^2$,
terms proportional to $m^4$ and $m^2{\rp}^2$ cancel so that we have
$$
V^{(1)} =  {\wt V}^{(1)}
	+  \1{64\pi ^2} {\big(}\2{\lambda }{\rp}^2{\big)}^2
	\times  O{\Big(} \muom, {\lambda {\rp}^2 \over m^2} {\Big)}
\eqn\eqONEPOTMDBLEii$$
where we denote the first term in Eq.~\eqONEPOTMDBLEi\
as ${\wt V}^{(1)}$,
which is just the potential of the low-energy effective theory.
This is an example of the `automatic' decoupling.

\Appendix{B}

We evaluate the fermion one-loop contribution
$\Lambda _{{\rm {\sss F}}}$
to the four-point vertex
$\Gamma _{\phi }^{(4)}
= -Z_{\lambda }\lambda
+ \Lambda _{{\rm {\sss B}}} + \Lambda _{{\rm {\sss F}}}$.
Generic one-loop integral can be reduced to the scalar-loop integral
	\refmark{\TV}
and expressed in terms of Spence function,
but we give another expression for $\Lambda _{{\rm {\sss F}}}$.
\foot{
The authors are grateful to T.~Kugo for discussions on this calculation.
}

The fermion one-loop contribution to the four-point vertex
$\Gamma _{\phi }^{(4)}$ is given by
$$\eqalign{
\Lambda _{{\rm {\sss F}}}(p_1,p_2,p_3,p_4)
	&= - {Ng^4 \over 16\pi ^2}\,I(p_1,p_2,p_3,p_4)
	 +~[~\hbox{permutation in}~p_2,~p_3,~p_4~] \ , \cr
I(p_1,p_2,p_3,p_4)
	&\equiv  16\pi ^2\,\mu _0^{2\epsilon } \LINT
	\Tr {\Big[}\,\1{~\ks(\ks+\ps)(\ks+\qs)(\ks+\rs)~}\,{\Big]} \cr
}\eqn\eqIDEF$$
where $p=p_1$, $q=p_1+p_2$ and $r=p_1+p_2+p_3=-p_4$.
This integral can be reduced to the scalar-loop integral
by carrying out the trace and making use of the identity
$2(k+a)(k+b) = (k+a)^2 + (k+b)^2 - (a-b)^2$
for any momenta $a$ and $b$.
As a result,
the integral $I$ decomposes into three parts, $I=I_2+I_3+I_4$,
where $I_n$ contains $n$ propagators.
We denote as $D_a\equiv (k+a)^2$.

The first integral $I_2$ is evaluated in cyclically symmetric way
by making a suitable shift in the loop momentum as
$$\eqalign{
I_2&=16\pi ^2\mu _0^{2\epsilon }\LINT~
	2{\Big[}\,\1{~D_pD_r~}+\1{~D_0D_q~}\,{\Big]} \cr
   &=16\pi ^2\mu _0^{2\epsilon }\LINT \1{~D_0D_{p_1+p_2}~}
	+~[~\hbox{cyclic in}~p_i~] \cr
   &= {\Big[}~\1{\ebar} + 2 + \ln{\mu _0^2 \over -(p_1+p_2)^2 }~{\Big]}
	~~~ +~[~\hbox{cyclic in}~p_i~]  \ . \cr
}\eqn\eqItwo$$
Similarly,
the second one $I_3$ is evaluated symmetrically as
$$\eqalignno{
I_3&=16\pi ^2\int {d^4k \over (2\pi )^4i}~
	{ - p_1^2 - p_2^2 + (p_1+p_2)^2 \over D_{-p_1}D_0D_{p_2} }
	 + [~\hbox{cyclic in}~p_i~] \cr
   &= J{\big(}-p_1^2,\, -p_2^2~;\, -(p_1+p_2)^2{\big)}
	+~[~\hbox{cyclic in}~p_i~] \ .
&\eqname\eqIthree \cr
}$$
We have introduced the function
$J(\xi ,\eta \,;\,\zeta )=J(\eta ,\xi \,;\,\zeta )$ defined
for $\xi ,\eta ,\zeta >0$ by
$$
J(\xi ,\eta ~;\,\zeta ) \equiv \int\nolimits_0^1 d\alpha ~
	{ \xi  + \eta  - \zeta  \over
	 \xi \,\alpha  + \eta \,(1-\alpha )
	- \zeta \,\alpha (1-\alpha ) }~
	\ln{ \xi \,\alpha  + \eta \,(1-\alpha ) \over
	\zeta \,\alpha (1-\alpha ) }
\eqn\eqJDEF$$
in terms of which we have useful formula
$$
16\pi ^2\int {d^4k \over (2\pi )^4i}~
{ - a^2 - b^2 + (a+b)^2 \over D_{-a}D_0D_{b} }
	= J{\big(} -a^2,-b^2~;\,-(a+b)^2\,{\big)} \ .
\eqn\eqJINT$$
Finally,
the $I_4$ is
$$
I_4=16\pi ^2\int {d^4k \over (2\pi )^4i}~
    { r^2(p-q)^2 + p^2(q-r)^2 - q^2(r-p)^2 \over D_0D_pD_qD_r}
\eqn\eqINTfour$$
which can also be expressed in terms of $J$
by making a conformal change
	\refmark{\TV}
of the integration variable
$k_{\mu }=(\mu ^2_0 /{\wb k}^2)\,{\wb k}_{\mu }$.
[$\mu _0$ is an arbitrary scale.]
Under the conformal transformation
${\wb a}_{\mu }\equiv (\mu ^2_0 /a^2)\,a_{\mu }$,
$(k+a)^2= (a^2/{\wb k}^2)({\wb k}+{\wb a})^2$,
we have
$$
\int {d^4k \over (2\pi )^4i}~\1{~D_0D_pD_qD_r~}
	= - { \mu _0^4 \over p^2q^2r^2 }
	  \int {d^4k \over (2\pi )^4i}~
	  \1{~D_{{\wb p}}D_{{\wb q}}D_{{\wb r}}~} \ .
$$
By applying the formula \eqJINT,
we evaluate the integral \eqINTfour\ as
$$\eqalignno{
I_4&=16\pi ^2\int {d^4k \over (2\pi )^4i}~
	{- ({\wb p}-{\wb q})^2
	 - ({\wb q}-{\wb r})^2
	 + ({\wb r}-{\wb p})^2
	\over D_{-({\wb q}-{\wb p})}D_0D_{{\wb r}-{\wb q}} } \cr
&=   J	{\Big(} -{\mu _0^4 \over p^2q^2}(p-q)^2\,,\,
		-{\mu _0^4 \over q^2r^2}(q-r)^2~ ;\,
		-{\mu _0^4 \over r^2p^2}(r-p)^2\,{\Big)} \cr
&= - \1{\,4\,}\,J{\big(}~
	p_1^2p_3^2\,,\, p_2^2p_4^2~ ;\, (p_1+p_2)^2(p_2+p_3)^2~{\big)}
	+~[~\hbox{cyclic in}~p_i~] \ .
&\eqname\eqIfour \cr
}$$
where we have used the definition \eqJDEF\
and the definition of $p$, $q$ and $r$ in the last equality
so that the cyclic symmetry becomes manifest.

Thus, the final form of the integral $I=I_2+I_3+I_4$ and
the fermion one-loop contribution $\Lambda _{{\rm {\sss F}}}$ are,
from Eqs.~\eqItwo, \eqIthree\ and \eqIfour,
$$\eqalignno{
I(p_i)&= I_0(p_1,p_2,p_3,p_4) + [~\hbox{cyclic in}~p_i~] \ ,
&\eqname\eqIFINALi \cr
\Lambda _{{\rm {\sss F}}}(p_i)
&= - {Ng^4 \over 16\pi ^2}~I_0(p_1,p_2,p_3,p_4)
 + [~\hbox{permutation in}~p_i~]
&\eqname\eqIFINALii \cr
}$$
with $I_0$ given in Eq.~\eqQUARTICfermionDEF.
At symmetric point,
$I_0$ reduces to the second term in Eq.~\eqZlambda\ with
$$
F   \equiv 2 + \ln \4{3}
	+ J{\big(}\,1,\,1\,;\3{4}\,{\big)}
	- \1{4} J{\big(}\,1,\,1\,;{16 \over 9}\,{\big)}
    = 3.02198\cdots
\eqn\eqFVALUE$$
where we numerically integrate the function
$J(\xi ,\eta \,;\,\zeta )\equiv
(\xi +\eta -\zeta )J_0(\xi ,\eta ,\zeta )$, \eqJDEF,
by using $E(\alpha )\equiv
\xi \alpha +\eta (1-\alpha )-\zeta \alpha (1-\alpha )$ and
$$\eqalign{
J_0(\xi ,\eta ,\zeta )
	=& {\xi +\eta  \over \xi \eta }
	  + \int\nolimits_0^1 d\alpha ~\1{E(\alpha )}
	    \ln{ \xi \,\alpha  + \eta \,(1-\alpha ) \over \zeta } \cr
	 &- \int\nolimits_0^1 d\alpha {\bigg\{}
		{\Big[}\,\1{E(\alpha )}-\1{E(0)}\,{\Big]}\ln \alpha
	  +	{\Big[}\,\1{E(\alpha )}-\1{E(1)}\,{\Big]}\ln(1-\alpha )
	 {\bigg\}} \ .\cr
}\eqn\eqJNUMERICAL$$

\refout
\endpage
\centerline{{\bf FIGURE CAPTIONS}}
\item{{\rm Fig.~1}}~
The `interpolating' functions
$K_{\psi }$ (solid line),
$K_g$ (dashed line) and
$K_{\lambda }$ (dot-dashed line).
\item{{\rm Fig.~2}}~
A typical behaviur of ${\wb m}^2$
in $\MDB$ (solid line), $\MS$ (dashed line)
and MD schemes (dot-dashed line).
All schemes as well as the full and effective theories in $\MS$ scheme
are matched at $\ln(\mu /m)=0$.
\item{{\rm Fig.~3}}~
The asymptotic behavior of the leading log potential
in $\MDB$ scheme (solid line), $\MS$ scheme (dotted line)
and MD scheme (dot-dashed line).
The inputs at $\mu =m$ are $g^2=0.55$, $\lambda =2.3$ (case (a)).
$V\sim {\rp}^4$ in the large $\rp$ region
and $V\sim {\rp}^2$ in the small $\rp$ region.
\item{{\rm Fig.~4}}
(a):~
The behavior of ${\wb \lambda }$ (solid line)
and  ${\wb g}^2$ (dashed line) in $\MDB$ scheme.
The input is moderate one: $g^2=0.55$, $\lambda =2.3$.
\nextline
(b):~
Same as Fig.~4~(a) for the extreme case: $g^2=0.5 $, $\lambda =2.5$.
${\wb \lambda }$ hits Landau singularity
in the high-energy region.
\nextline
(c):~
Same as Fig.~4~(a) for the extreme case: $g^2=0.7$, $\lambda =2.0$.
${\wb \lambda }$ becomes negative to cause the vacuum instability.
\item{{\rm Fig.~5}}
(a):~
The `character' $\chi _{{\MDS}}$ (solid line),
$\chi _{{\MSS}}$ (dashed line) and
$\chi _{{\rm {\sss MD}}}$ (dot-dashed line)
for the moderate case (a).
The large improvement is obtained in asymptotic regions.
\nextline
(b):~
Same as Fig.~5~(a) for the extreme case (b).
Due to $K_{\lambda }$,
sizable scheme dependence is observed near Landau singularity.
\nextline
(c):~
Same as Fig.~5~(a) for the extreme case (c).
Since the mass term dominates the quartic term in $V$,
sizable difference from $\MS$ and $\MDB$ is observed
in MD scheme.

\draw{\epsfbox{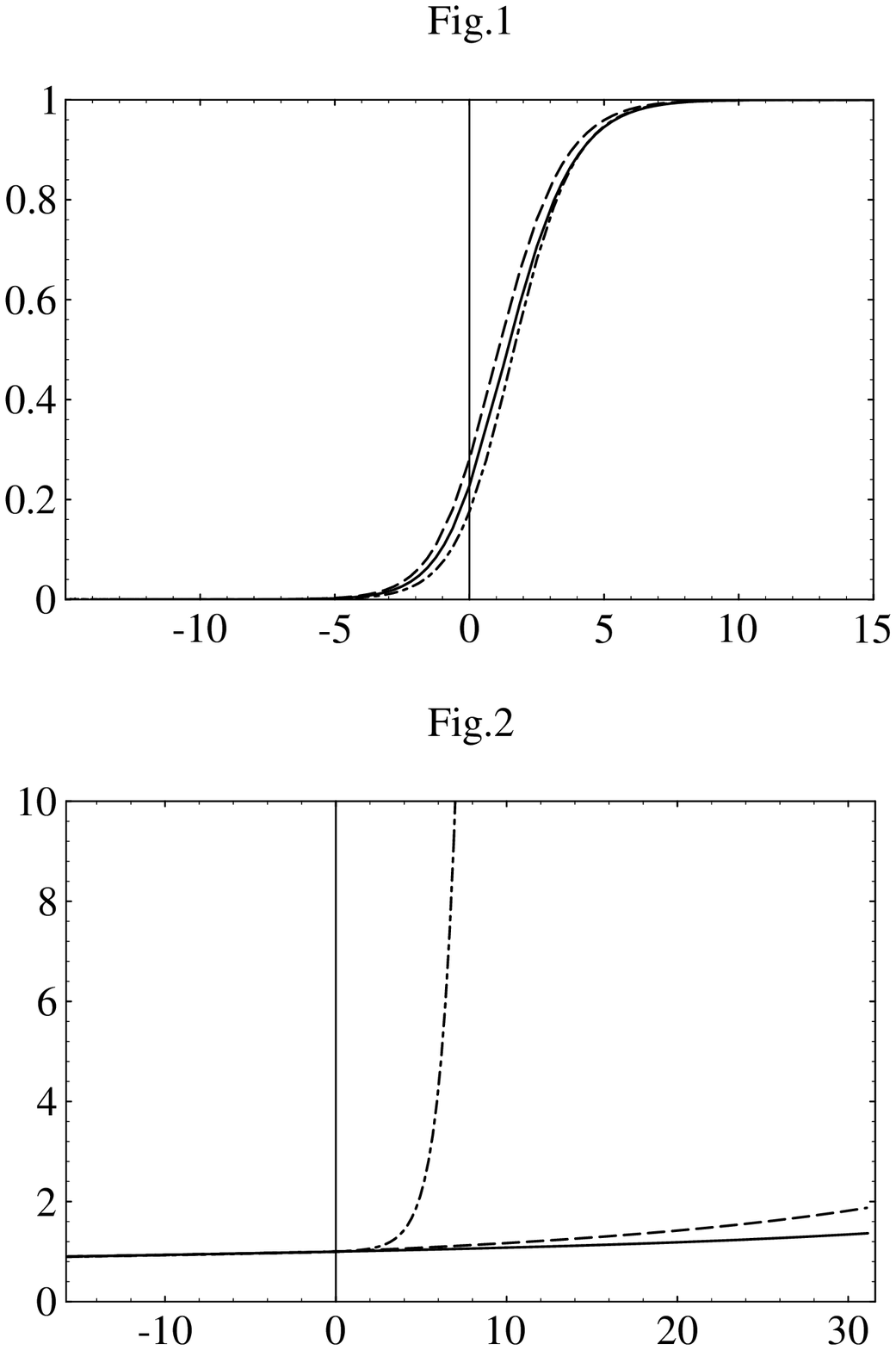}}
\draw{\epsfbox{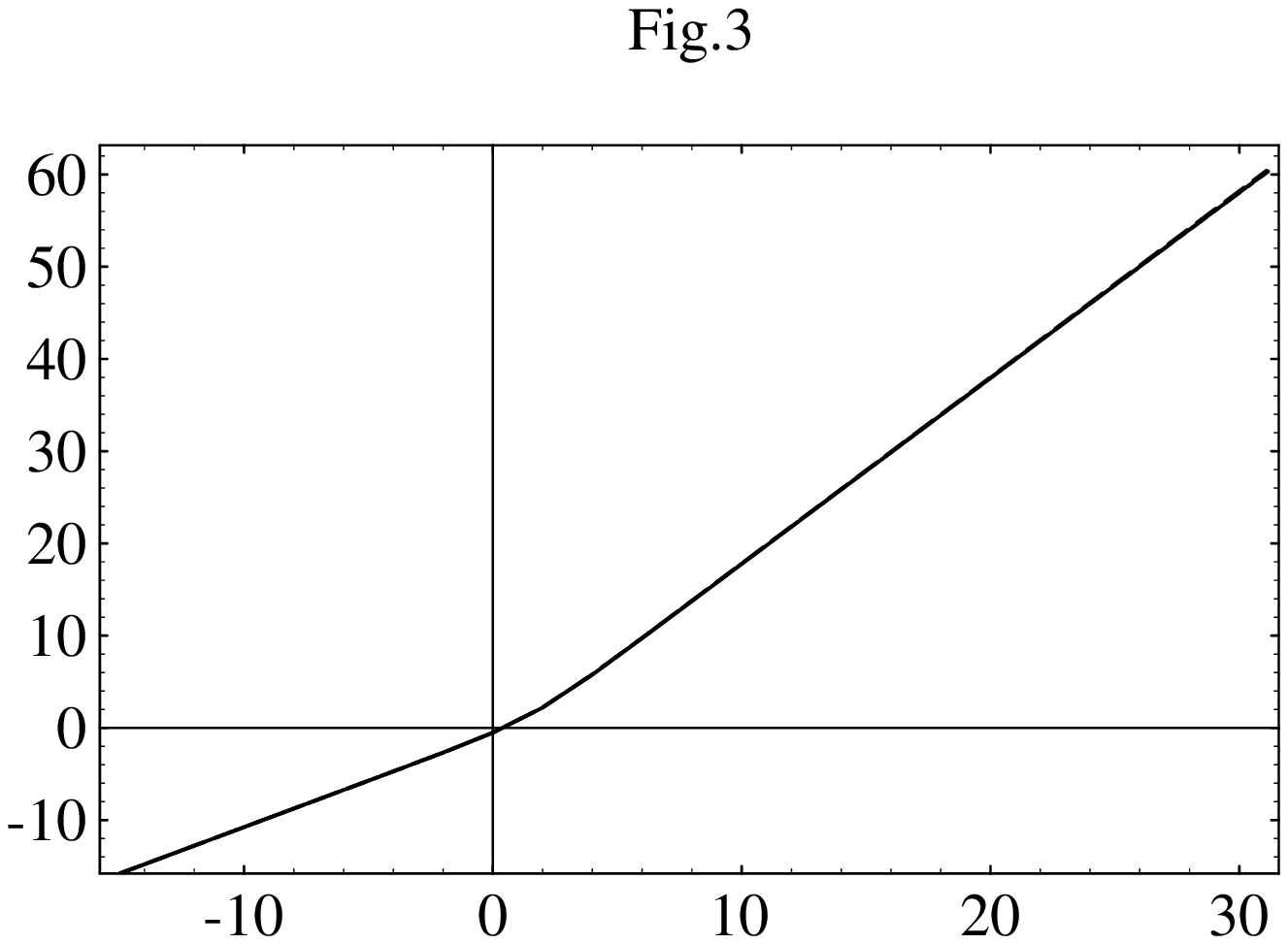}}
\draw{\epsfbox{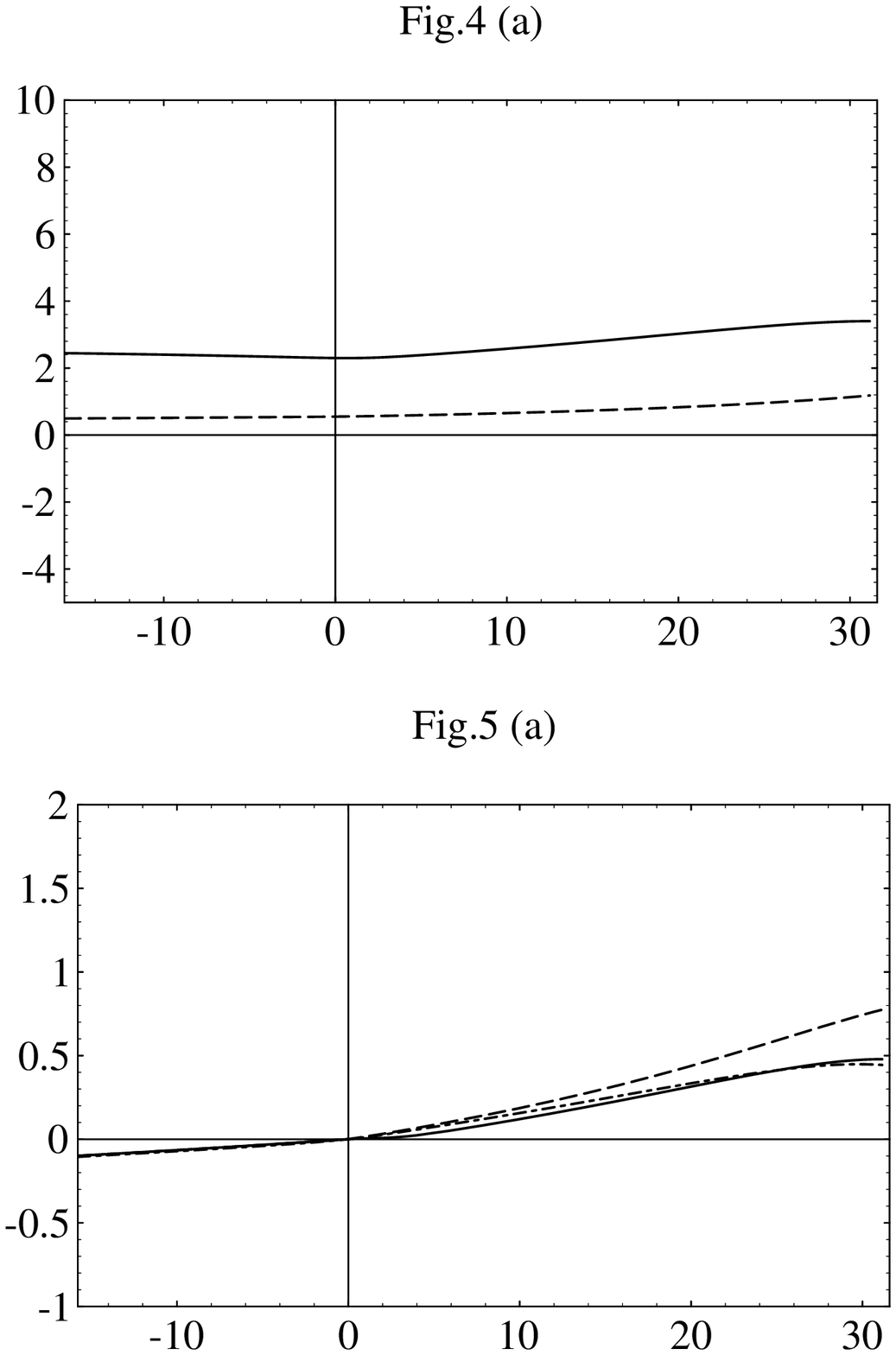}}
\draw{\epsfbox{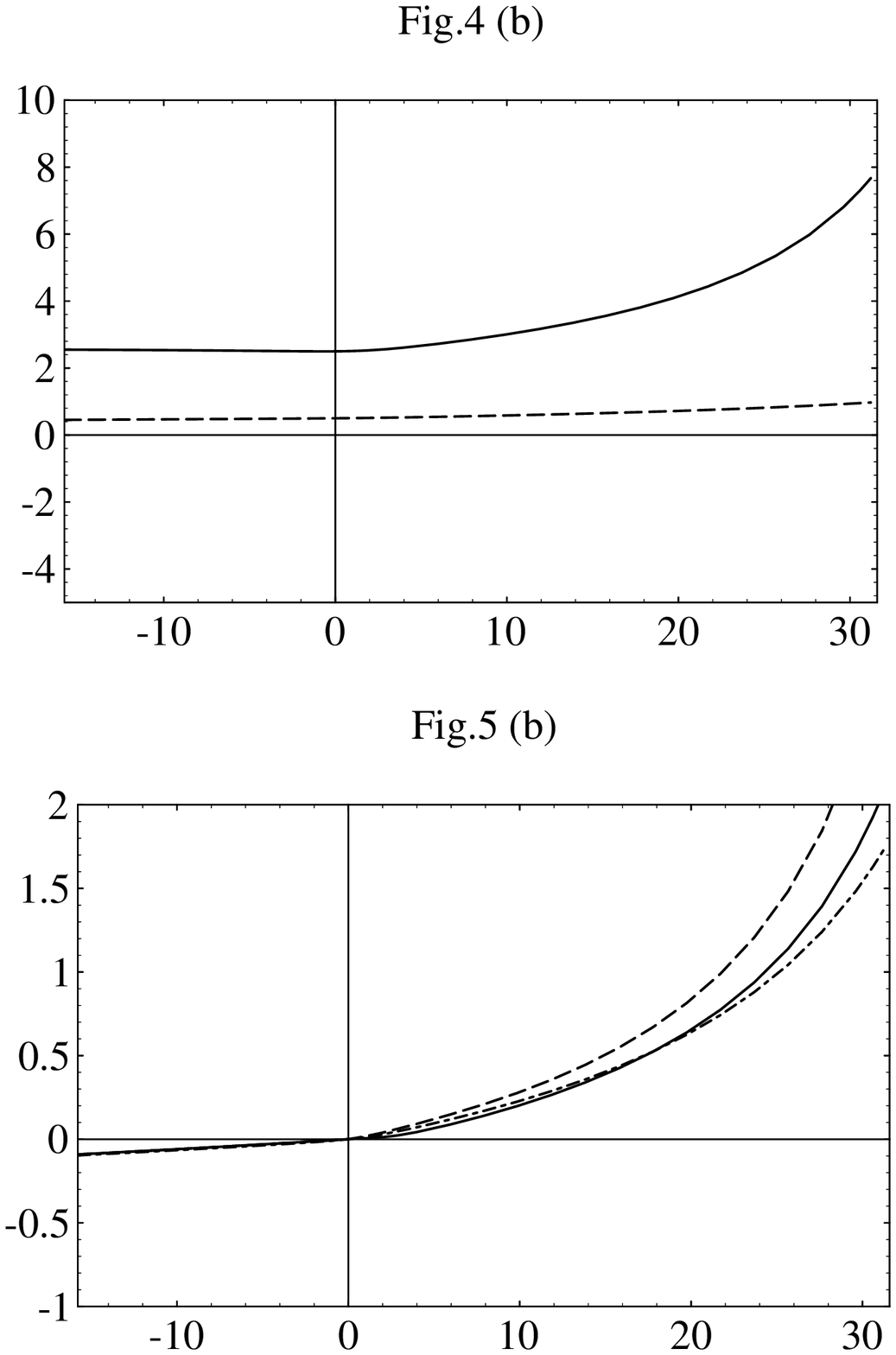}}
\draw{\epsfbox{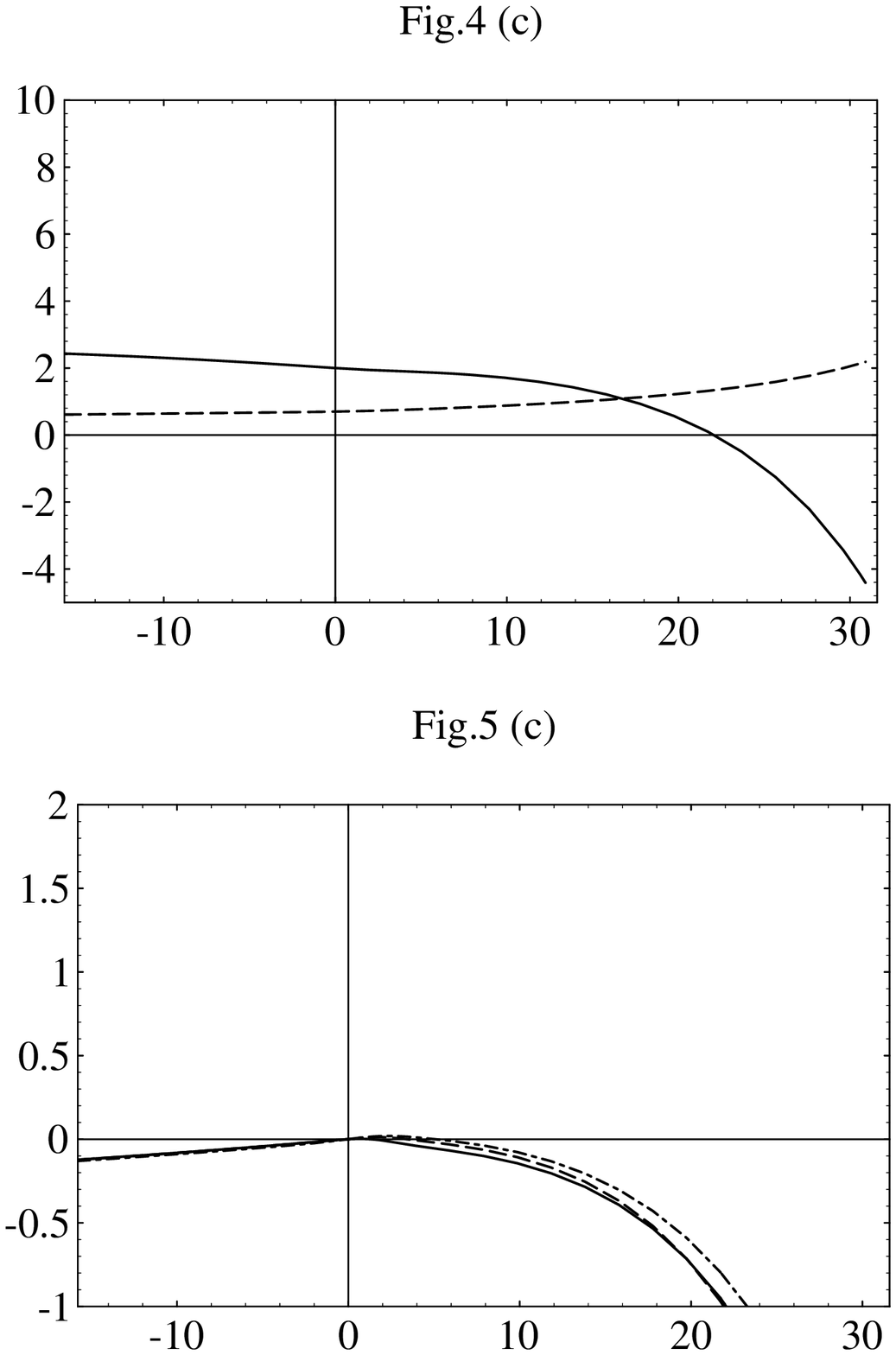}}

\bye\bye